\documentclass[10pt,journal]{IEEEtran}
\usepackage{amsfonts}
\usepackage[pagewise,switch]{lineno}
\usepackage{booktabs}
 \usepackage{multirow}
\usepackage{times}
\usepackage[ruled,linesnumbered,vlined]{algorithm2e}
\usepackage{cite}
\usepackage{indentfirst}
\usepackage{amsmath}
\usepackage{multirow}
\usepackage{float}
\usepackage{soul}
\usepackage{color}
\usepackage{subfigure}
\usepackage{graphicx}
\usepackage{caption}
\usepackage{multirow}
\usepackage{amsthm}
\usepackage{ulem}
\begin{document}

\title{EmoSense:  Computational Intelligence Driven Emotion Sensing via Wireless Channel Data}

\author{Yu~Gu ,~\IEEEmembership{Senior Member,~IEEE,}
        Yantong~Wang,
        Tao~Liu,
        Yusheng~Ji,~\IEEEmembership{Senior Member,~IEEE,}
        Zhi~Liu,~\IEEEmembership{Member,~IEEE,}
        Peng~Li,
        Xiaoyan~Wang,
        Xin~An,
        and Fuji~Ren,~\IEEEmembership{Senior Member,~IEEE,}
\IEEEcompsocitemizethanks{
\IEEEcompsocthanksitem Y. Gu (co-corresponding author), Y. Wang, T. Liu and X. An are with School of Computer and Information, Hefei University of Technology, China.
E-mail: yugu.bruce@ieee.org, \{wangyantong912 and LTao\}@mail.hfut.edu.cn
\IEEEcompsocthanksitem Y. Ji is with National Institute of Informatics, Japan.
E-mail: kei@nii.ac.jp
\IEEEcompsocthanksitem Z. Liu is with Shizuoka University, Japan
E-mail: liu@ieee.org
\IEEEcompsocthanksitem X. Wang is with  Ibaraki University, Japan.
E-mail:  xiaoyan.wang.shawn@vc.ibaraki.ac.jp
\IEEEcompsocthanksitem P. Li is with  University of Aizu, Japan.
E-mail:  pengli@u-aizu.ac.jp
\IEEEcompsocthanksitem  F. Ren (co-corresponding author) is with University of Tokushima, Japan.
E-mail: ren@is.tokushima-u.ac.jp
}}

\maketitle

\markboth{IEEE Transactions on Emerging Topics in Computational Intelligence}
{Gu \MakeLowercase{\textit{et al.}}: EmoSense: Ubiquitous Data-driven Emotion Sensing via Commodity WiFi Infrastructures}

\begin{abstract}
Emotion is well-recognized as a distinguished symbol of human beings, and it plays  a crucial role in our daily lives. Existing vision-based or sensor-based solutions are either obstructive to use or rely on specialized hardware, hindering their applicability. This paper introduces EmoSense, a first-of-its-kind wireless emotion sensing system driven by computational intelligence. The basic methodology is to explore the physical expression of emotions from wireless channel response via data mining.  The design and implementation of EmoSense {face} two major challenges: extracting physical expression from wireless channel data and recovering emotion from the corresponding physical expression. For the former, we present a Fresnel zone based theoretical model depicting the fingerprint of  the physical expression on channel response. For the latter, we design an efficient computational intelligence driven mechanism to recognize emotion from the corresponding fingerprints. We prototyped EmoSense on the commodity WiFi infrastructure and compared it with main-stream sensor-based and vision-based approaches in the real-world scenario. The numerical study over $3360$ cases confirms that EmoSense  achieves a comparable performance to the vision-based and sensor-based rivals under different scenarios. EmoSense only leverages the low-cost and prevalent WiFi infrastructures and thus constitutes a tempting solution for emotion sensing.

\end{abstract}

\section{Introduction}
Emotion is a significant feature of human beings. It is also the key to {interpreting} implicit messages in human interaction \cite{Fragopanagos2002Emotion}. Though humans seem to be born with innate emotional capabilities, it is not a natural gift for the computers. Therefore, emotion sensing becomes an emerging topic for the human-machine interaction with various tempting applications like elder emotion companion \cite{HANJing2015EmoElder} and autism treatment \cite{El2010Affective}.

Emotion, as a complicated psychological state, usually exhibits both external signature like physical expression, and internal signature like physiological signal. Accordingly, current emotion sensing solutions can be divided into two categories, i.e., vision-based \cite{Ioannou2005Emotion,Wang2017Speech}  and sensor-based \cite{Jenke2017Feature,Katsigiannis2017DREAMER}. The former focuses on capturing external signature for emotion recognition, e.g., facial expression \cite{Ioannou2005Emotion} or body gestures \cite{Glowinski2008Technique}. The latter concentrates on detecting internal signature for recovering emotions, e.g.,  electroencephalogram (EEG) signals for evaluating inner emotional status \cite{Jenke2017Feature}.

The last few decades have witnessed solid research progresses in emotion sensing achieved by the above two mainstream solutions. However, they still have some fundamental yet unsolved issues. For instance, current systems are usually built on specialized hardware, making their availability a prominent problem. Also, they are normally constrained by physical and environmental conditions such as illumination and line-of-sight (LOS) dependence, leading to the reliability issue. Last but not least, they could be considered as offensive since people usually dislike physical contact (sensors) or being monitored (cameras). Hence people are seeking for possible alternatives to innovate the conventional approaches by asking the following question:

\emph{How can we construct an emotion sensing system that (1) effectively recognizes emotions without any specialized devices, (2) robustly works under different circumstances like site, target, illumination condition, and (3) continuously monitors the area of interest without privacy concern?}

In this paper, we introduce EmoSense, a first-of-its-kind wireless emotion sensing system that leverages channel response from off-the-shelf WiFi devices. EmoSense has three major advantages compared to its vision-based and sensor-based rivals. Firstly, it does not {rely} on specialized hardware since the low-cost WiFi infrastructure is pervasive nowadays. Secondly, it is robust since characterizing
the channel response with Fresnel zones waives the environmental dependence. Lastly, it is contactless and free of privacy concern since the WiFi signal is unnoticeable for users.

EmoSense explores body gesture that contains rich mood expressions for emotion recognition. The key idea is that human body gesture affects wireless signal via the shadowing and multi-path effects. Such effects usually form unique patterns or fingerprints in the temporal-frequency domain for different gestures. EmoSense leverages the gesture fingerprint to recover the corresponding emotions. Its design and implementation faces two challenges, i.e.,

\begin{enumerate}
 \item  How to identify the body gesture through its fingerprint on the wireless signal?
   \item How to recognize emotions from its corresponding body gestures (physical expressions)?
\end{enumerate}

The first challenge corresponds to enhancing and extracting the fingerprint of body gestures (sometimes very minor and brief) on wireless signal in terms of channel response. To this end, we propose a look-up method based on Fresnel zones to ensure a quick and efficient experimental setup to capture fine-grained gestures.

The second challenge equals to matching the gestures to the corresponding emotions. It essentially converges to a typical data mining problem that can be be solved by computational intelligence. Therefore, we design a computational intelligence driven architecture to explore both temporal and frequency features from signal fingerprints to recover emotions.

We prototype EmoSense with low-cost off-the-shelf WiFi devices and evaluate its performance in the real environments.  We also realize two traditional vision-based and sensor-based systems  for the comparative study. All three systems focus on the external signature of emotion. We recruit 14 subjects with no act training and ask them to evoke four emotions (happy, sad, anger and fear) through audiovisual stimulations, i.e., watching video clips or listening to music. During the experiment, the vision-based system is capturing the facial expressions, while EmoSense and the sensor-based system keep monitoring the simultaneous body gestures, respectively.

The comparative study over $3360$ cases suggests that EmoSense \emph{1)} achieves a comparable performance to the vision-based and sensor-based rivals under different scenarios with a classic k-Nearest Neighbor (kNN) classifier,  \emph{2)} works robustly since the impact of external circumstances like site, target, illumination, and line-of-sight is limited, and \emph{3)} is unobtrusive since no subject reports privacy or comfort complaints as in the vision-based and sensor-based rivals. Furthermore, we report several interesting investigations. For instance, the empirical result confirms that the physical expression of emotions is person-dependent. In other words, different people have different habits of expressing their moods.

Our contributions can be summarized as follows:
\begin{enumerate}
\item We design a Fresnel zone based model to characterize the physical expression of emotion on wireless channel data and provide a look-up method to enhance the fingerprints by adjusting the experimental settings.

\item We devise a computational intelligence driven scheme to effectively extract key features and efficiently recognize emotion from the CSI amplitude data.

\item We realize EmoSense, a first-of-its-kind WiFi-based emotion sensing system, on the commodity WiFi devices. EmoSense has been evaluated with a vision-based system and a sensor-based system in real environments. The experimental results not only confirm its effectiveness, but also reveal several inspiring observations.

\end{enumerate}

The  rest of this paper is organized as follows: we introduce a literature review in the next section, following by some preliminaries that inspire the design of EmoSense in section \ref{sect:Pre}. In section \ref{sect:SysDe}, we present the detailed design of EmoSense. Then, we evaluate EmoSense in real scenarios and explain the experimental results in section \ref{sect:PerEva}. Finally, we conclude our work and outline some possible extensions in section \ref{sect:Conclusion}.

\section{Related Works}
{This work involves two topics, i.e., affective computing and WiFi-based gesture recognition. We will introduce the related research for both topics in this section.}
\subsection{{Affective Computing}}
Two decades ago, Marvin Minsky \cite{Minsky1987The} raised the famous argument: ``The question is not whether intelligent machines can have any emotions, but whether machines can be intelligent without emotion?''. Since then, affective computing, which intends to endow computers with the ability to timely sense users' moods and to intelligently respond them, becomes a rising star of computer science and attracts tons of attention from both industry and academia \cite{Picard1997Affective}. One essential methodology has been laid down, i.e., exploring emotion via its expressive modalities such as audiovisual clues, textual input, physiological signals, and body gestures.

\emph{\textbf{Audiovisual-based:}} In the daily life, voice and facial expressions embody most of our emotional elements. As a result, 95\% of our current research on emotion recognition relies on the facial expression as stimuli. For instance, Saste and Jagdale \cite{Saste2017Emotion} designed a system recognizing emotion in speech that is irrelevant to languages. The system can be embedded in Automated Teller Machines (ATMs) for the safety purpose. Recently, Liu \emph{et al.} \cite{Liu2017A} proposed FEER-HRI, an online system that can not only recognize emotion during communication  between human and robots via facial expressions, but also generate the corresponding emotion on robot for better interaction.

\emph{\textbf{Textual-based:}} 80\% of our historical knowledge has been preserved in text. Nowadays, as online social media like Facebook, Wechat, and Twitter become indispensable in the modern society, it is a tempting way of exploring rich textual social information for implicit emotions.

Generally speaking, emotional words are commonly seen in the text documents no matter in which language they are written. Shivhare \emph{et al.} \cite{Shivhare2015EmotionFinder} leveraged the emotion word ontology and classified them into different emotion levels that have different scores.  Then, emotion of the input text can be determined by mapping the sum of emotion scores in the text to certain emotion categories.

It is a common sense that only words are far from enough to infer the inherent complex emotions in the text. Therefore, the syntactic and semantic structure of the text are frequently used. For instance, Shaheen \emph{et al.} \cite{Shaheen2015Emotion} proposed ERR, a novel method leveraging the syntactic and semantic structure of the input  English sentence and extracting emotional information for emotion recognition.

As the research on both classes moves forwards, one prime concern attracts more and more attention, i.e., both audiovisual clues and textual {inputs} are vulnerable to the intentional emotion induction and masking, because they only represent artificial emotions that are not direct and could be tuned. To this end, there comes a new upsurge on directed reflections of emotion based on physiological signals and body gestures.

\emph{\textbf{Physiological-signal-based:}} The most commonly-used physiological signals are heart beating rate, breathing rate, blood pressure, and skin conductance. Usually, those signals are obtained by contact or invasive sensors. A comprehensive survey on emotion recognition via physiological sensors is presented in \cite{Wioleta2013Using}.

Recently, researchers tend to the ubiquitous wireless signals for physiological measurements in a non-contact way. For example, Zhao \emph{et al.} \cite{Zhao2016Emotion} designed EQ-radio, a one-of-its-kind emotion recognition system using wireless signals to physiological info such as the heart beating rate and the breathing rate.

\emph{\textbf{Gesture-based:}} It is reported that human gestures also possess ample emotional elements that are not fully explored yet. Lv  \emph{et al.} \cite{Lv2008Emotion} were among the first to design a gesture-based emotion recognition system through analyzing the typing sequence on a keyboard. However, Pusara \emph{et al. } \cite{Pusara2004User} found that the mouse movement alone is not enough for emotion recognition since it includes too little emotional info. To this end, body movements that are rich in emotion have been explored recently. Glowinski \cite{Glowinski2008Technique} \emph{et al.} are among the first to compute emotion by analyzing body movements through the off-the-shelf cameras. Piana \emph{et al.} \cite{Piana2014Real} pushed the research further by utilizing the kincet device to extract postures, physical features, and movement trends from the three-dimensional skeleton of the human body.

\subsection{{WiFi-based Gesture Recognition}}
{It is well-known that human beings interfere the wireless signals due to multi-path and fading effects} \cite{perera2014context}. {But only until recently such interferences have been explored for gesture recognition} \cite{Gu16IoT}.

{The cost-effective WiFi infrastructure is widely accessible nowadays.The most commonly-used indictor for the channel response of WiFi is the Received Signal Strength (RSS), a coarse-grained power feature summed over all propagation paths. Sigg \emph{et al.} were among the first to explore RSS for recognizing hand gesture}\cite{sigg2014telepathic}. {Later, Gu \emph{et al.} showed that RSS is also applicable for the whole-body gestures} \cite{Gu16IoT}.

{RSS is handy, but incapable when dealing with the multi-path effect. Therefore, Channel State Info (CSI), which characterizes the wireless signals with the frequency, amplitude (energy feature) and phase information, soon comes in as a better alternative} \cite{Yang2013From}. {Zeng \emph{et al.} use CSI in recognizing hand {gestures} and achieve better performance} \cite{zeng2014your}. {Soon CSI has been explored for fine-grained {gestures} such as mouth movements} \cite{wang2014we} {and keystrokes} \cite{Ali:2015:WiKey,Chen:2015:TKU}.

Though the gesture-based affective computing becomes more and more popular nowadays,  traditional solutions relying on vision and wearable sensors embody several crucial demerits such as the availability, reliability and privacy issues. To this end, we present a early version of EmoSense \cite{Gu2018ICC} to demonstrate the feasibility of exploring channel response for emotion recognition. In this paper, we push the research much further by elaborating the system design with computational intelligence. The enhanced system has been extensively evaluated with its vision-based and sensor-based rivals in real environments. The result shows that EmoSense achieves quite competitive performance.

\section{Preliminaries}
\label{sect:Pre}
In this part, we will first introduce the basic concepts of wireless channel data, where the fingerprints of human motion and emotion are hidden. Then, we will build a prototype to conduct a pilot experiment studying how the physical expression of emotion affects the signals.

\subsection{Overview of Wireless Channel Data}
EmoSense is driven by wireless channel data, where there exist two options provided by the physical layer (PHY), i.e., Received Signal Strength (RSS) and Channel State Info (CSI). The former is coarse-grained and represents the total received power level at the receiver, while the latter is fine-grained and describes signal attenuation from both time and frequency domains. RSS is usually obtained as follows \cite{Yang2013From},

\begin{equation}
\label{equ:RSS}
RSS=10 \log_{2}{({\Vert H \Vert}^2)} ,
\end{equation}
where $H$=$\sum_{k=1}^N \Vert H_k \Vert e^{j\theta_k}$. $\Vert H_k \Vert$ and $\theta_k$ represent the amplitude and phase on the $k$-th signal propagation path, respectively.

Equation (\ref{equ:RSS}) implies why RSS is considered to be a coarse-grained indicator because it only characterizes the total received power over all possible paths. In other words, RSS is unable to process the multi-path effect.

To this end, there is a recent trend of exploring CSI, a fine-grained indicator, to extract  multi-path channel features for motion detection \cite{Gu16IoT,Gu17IoT}. More specifically, current WiFi protocols are based on the Orthogonal Frequency Division Multiplexing  (OFDM) system, where $H(f,t)$ is a complex value of channel frequency response (CFR) in terms of CSI. It describes channel performance with the amplitude and phase information for the subcarrier frequency $f$ measured at time $t$. It is usually formulated as follows \cite{Tse2009Fundamentals}.

\begin{eqnarray}
\label{equ:CSI}
H(f,t)=\sum_{k=1}^N h_k(f,t) e^{-j\theta_k(f,t)},
\end{eqnarray}
where $h_k$ represents the amplitude and $e^{\theta_k(f,t)}$ indicates the phase shift on the $k$-th path caused by the propagation delay.

As in our previous WiFi-based PAWS \cite{Gu16IoT} and MoSense \cite{Gu17IoT} systems, EmoSense also employs the fine-grained CSI for the channel data.

\begin{figure}
\centering
\includegraphics[width=\columnwidth]{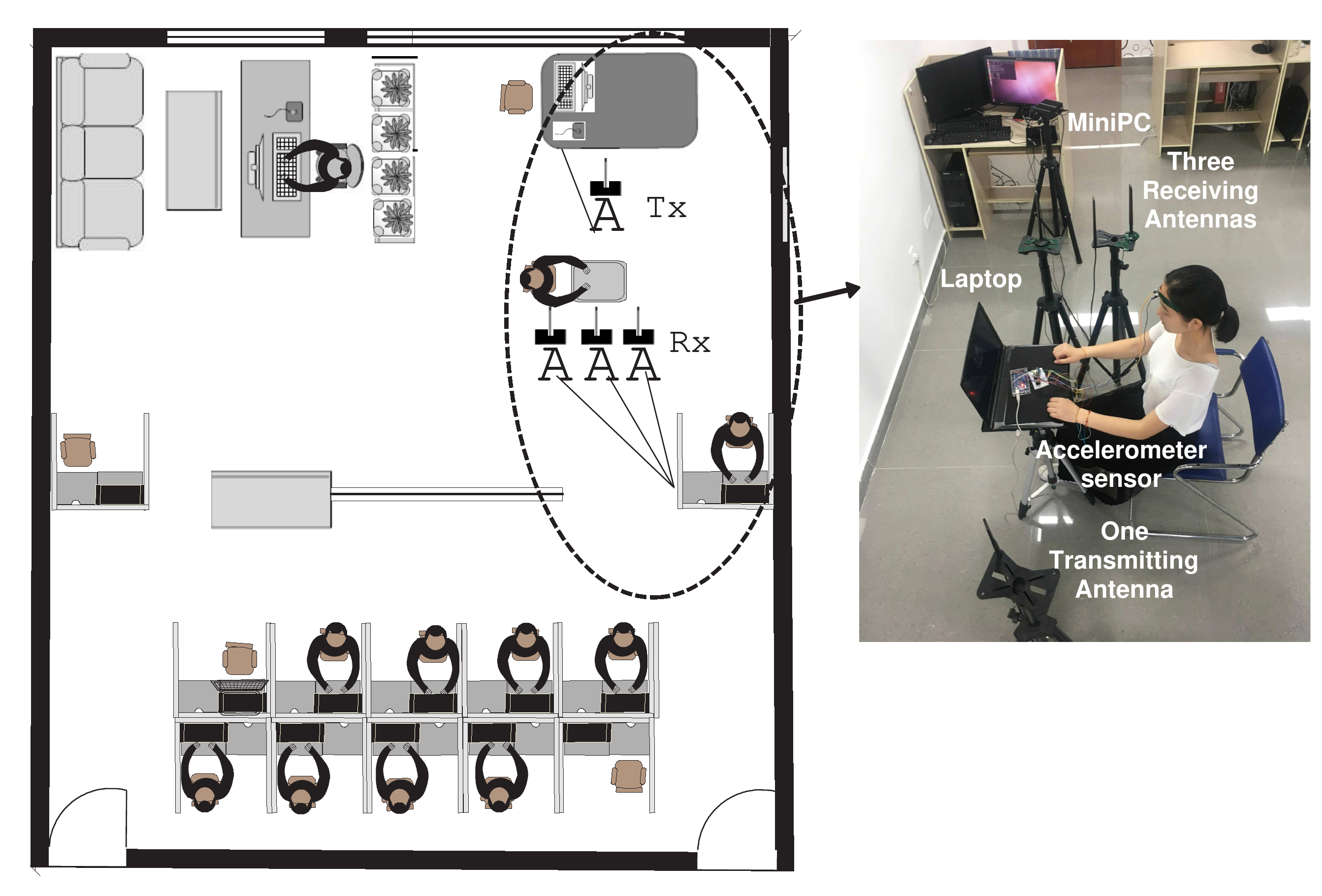}
\caption{Our prototype system.}
\label{fig:prototype}
\end{figure}

\begin{figure*}[t]
\centering
\hspace{-6ex}
\subfigure[The body gesture of emotion indeed interferes with channel response]{
\begin{minipage}[t]{0.33\textwidth}
\includegraphics[width=1.1\columnwidth]{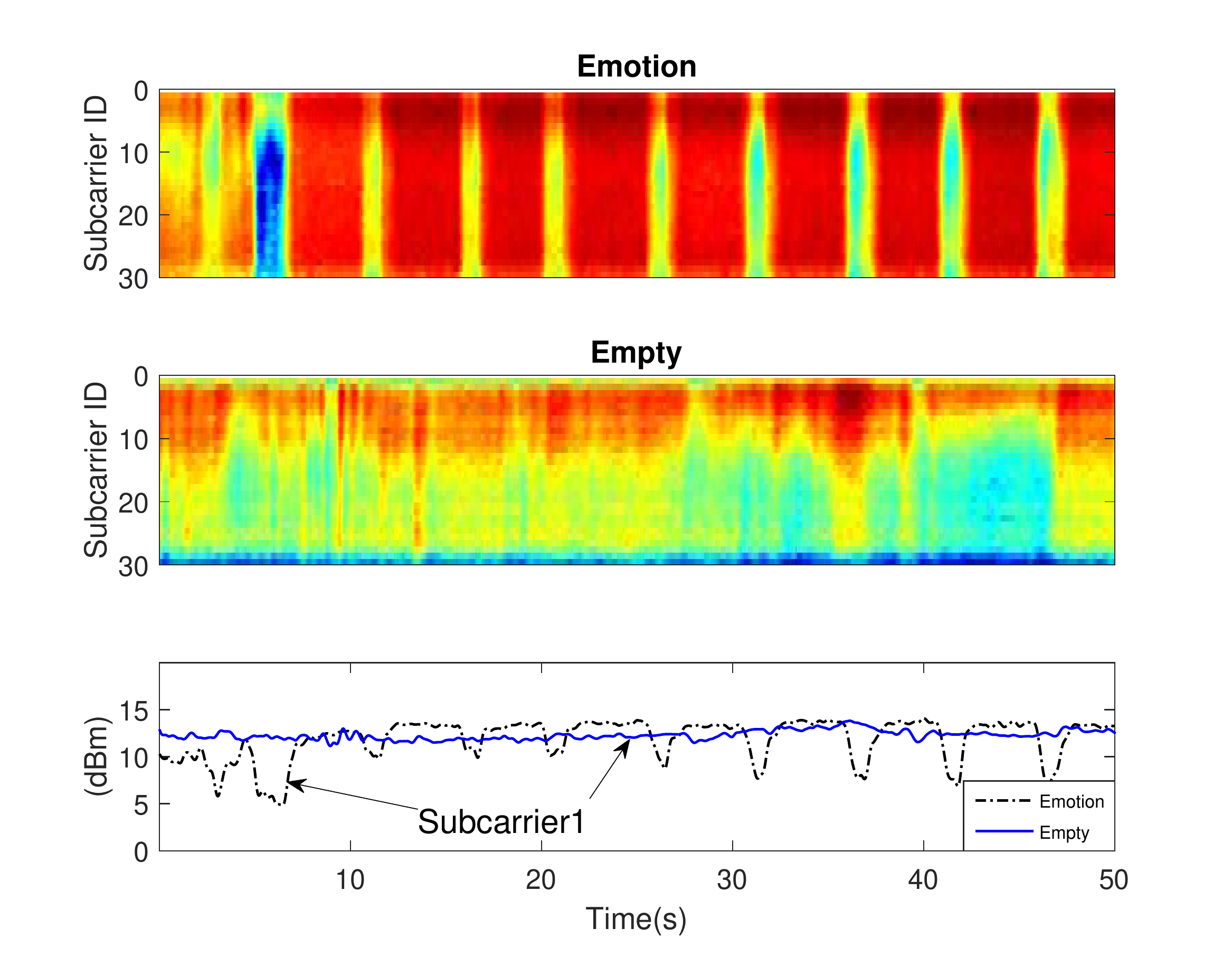}
\label{fig:MotionAffectsCR}
\end{minipage}
}
\hspace{-1ex}
\subfigure[The layout of antennas is critical for capturing fine-grained gesture fingerprints]{
\begin{minipage}[t]{0.33\textwidth}
\centering
\includegraphics[width=1.1\columnwidth]{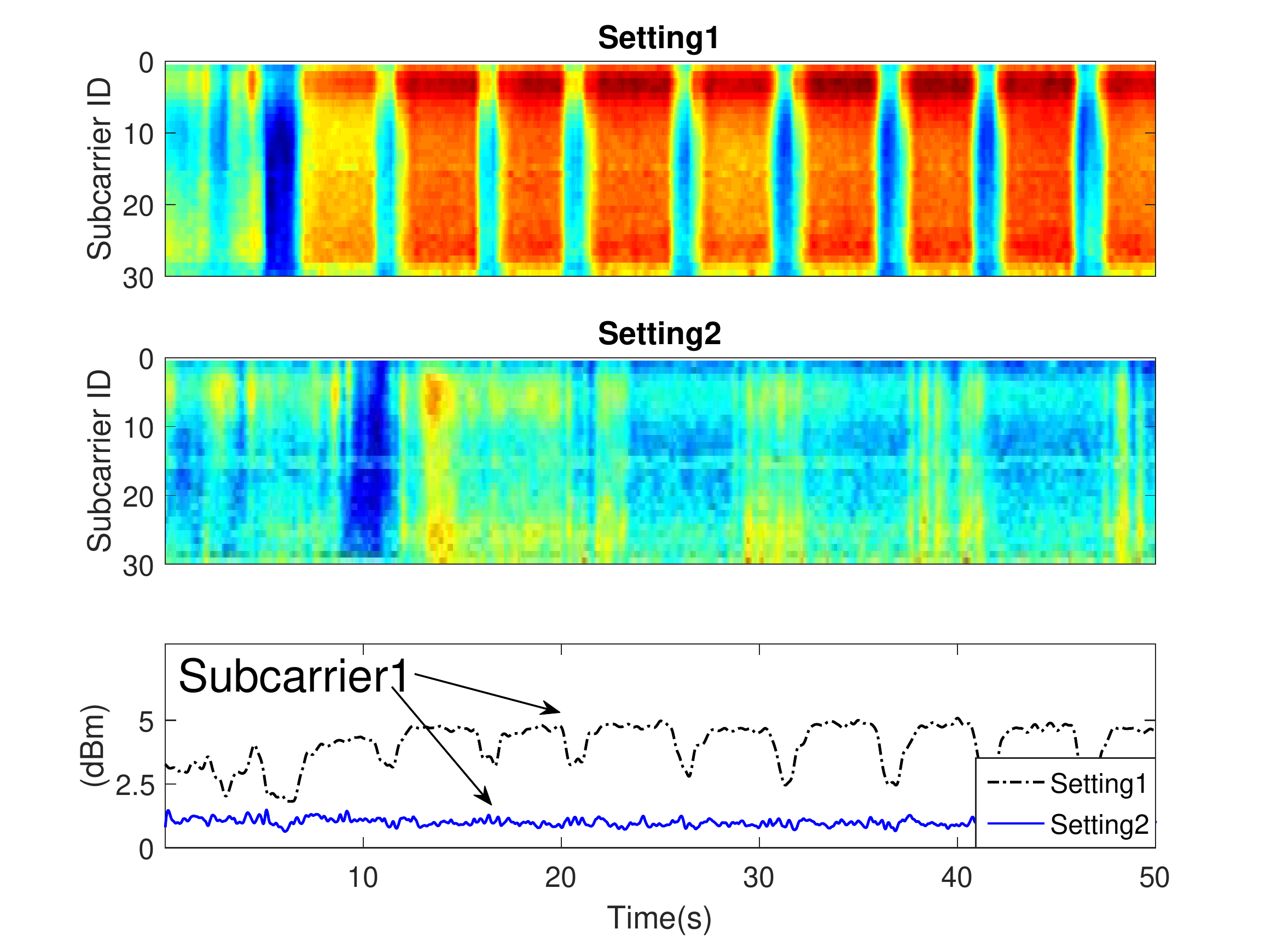}
\label{fig:SettingAffectsResult}
\end{minipage}
}
\hspace{-1ex}
\subfigure[The body gesture of emotion is person-dependent]{
\centering
\begin{minipage}[t]{0.33\textwidth}
\centering
\includegraphics[width=1.1\columnwidth]{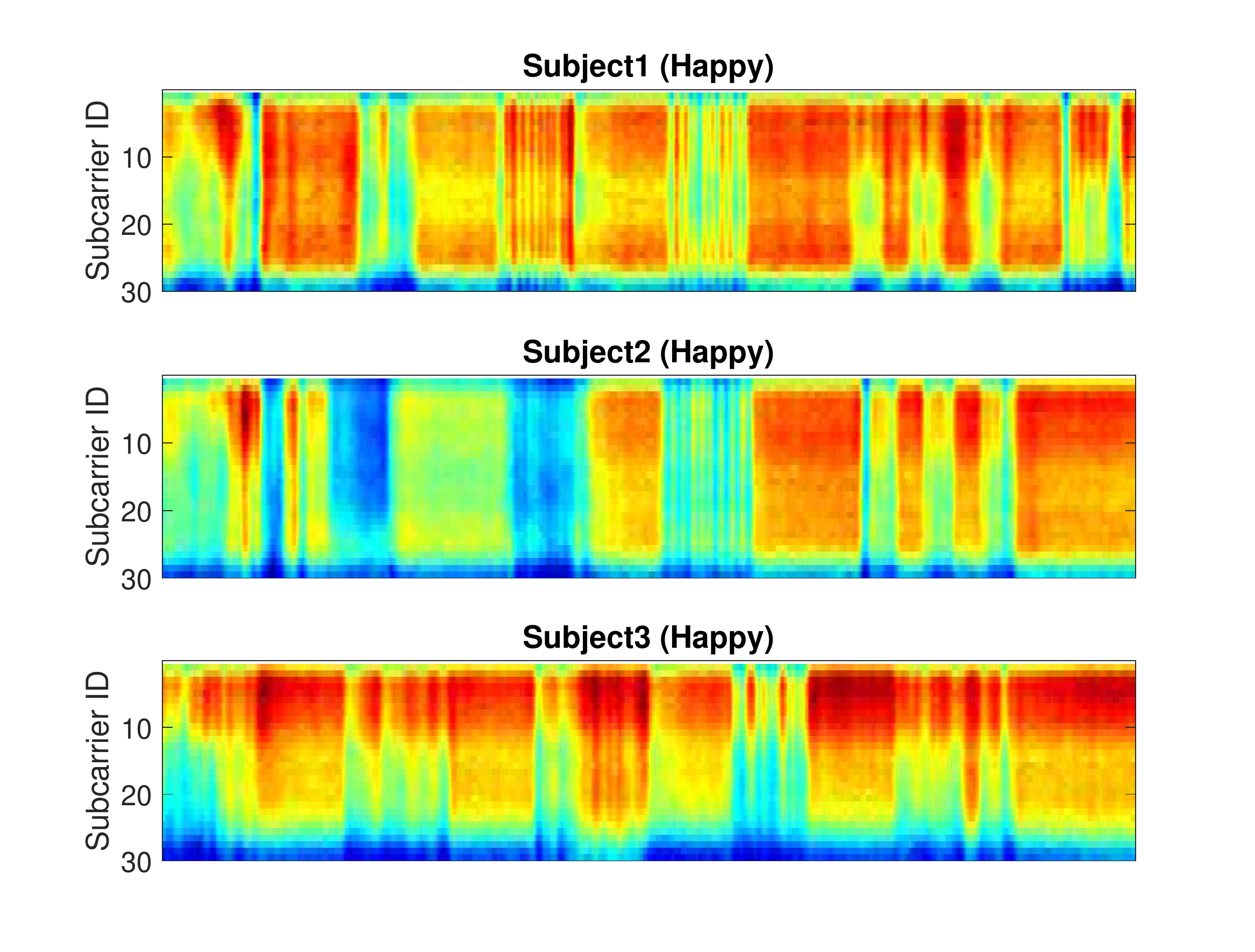}
\label{fig:EmotionIsPersonDependent}
\end{minipage}
}
\caption{Key observations inspiring the design of EmoSense in preliminary experiments.}
\label{fig:Prelimary}
\end{figure*}

\subsection{Preliminary Experiments}

In order to examine how the physical expression of emotion affects the signal, a pilot experiment is conducted as follows.

\noindent \textbf{[Prototype].}
Our prototype comprises two commodity MiniPCs, mounted with Intel Network Interface Controller (NIC) 5300 ($5$GHz)  (See Fig \ref{fig:prototype}), one of which is the sender with one external antenna, while the other one is the receiver with three antennas. These antennas are fixed on tripods. The sampling rate is 100Hz.

\noindent \textbf{[Participants].} 14 participants (5 females), aging from 21 to 26, are involved in the experiments. None of them has received any acting training to guarantee the natural expressions of {emotions}.

\noindent \textbf{[Environment].} The experiments were carried out in a $7\times 10$ $m^2$ office room, which contains some office furniture, such as couches, chairs, office tables and book shelves. During the experiments, some students are in their spots in the same room.

\noindent \textbf{[Emotions].} Four emotions are distinguished, i.e., happiness, sadness, anger and fear. Different audiovisual stimulations, e.g. watching video clips or listening to the music, are used to arouse  different emotions of the participants, and they are asked to perform accordingly.

Through the above experiments, careful observations are made.

\noindent \textbf{\emph{1) The body gesture of emotion indeed interferes with channel response:}}
The channel response data is indeed affected by the physical expressions of emotions, for example in Fig.\ref{fig:MotionAffectsCR}. It is visualized in respect of amplitude of one subject's physical expressions of happiness.  The subject was asked to watch a one-minute comedy, with body-movements like clapping and leaning back and forth and laughter. The channel data captured during the experiment is presented in the top figure, and that of the empty state (i.e. free of human intervention) is shown in the middle figure for comparison. The results show that body-movements (physical expressions) significantly affect the channel data. In other words, the channel response data is indeed affected by the physical expressions of emotions.

\noindent \textbf{\emph{2) The layout of antennas is critical for capturing fine-grained gesture fingerprints:}} The channel response data on physical expressions of emotions subject to different experimental settings, as a little adjustment of which can significantly affect the fingerprint on the channel data. For example, two different settings were used to record the same physical expression of one participant in Fig \ref{fig:SettingAffectsResult}. The transmitting antenna in Setting 1 is moved 20cm closer to the participant in Setting 2. This minor adjustment produces significantly different results. The channel response data captured in Setting 1 exhibits a much stronger fingerprint of physical expressions; however, it is not clearly shown in Setting 2. To clarify this point, both settings are compared in terms of subcarrier \#1 in the bottom figure of Fig. \ref{fig:SettingAffectsResult}. The previous phenomenon is testified again. The signal attenuation under the influence of the physical expressions is clearly recorded by subcarrier 1 in Setting 1, while it is barely shown by the same subcarrier in Setting 2. This observation repeatedly appears in our experiments, rendering us to wonder the causes of this phenomenon and the ways to enhance it.

\begin{figure}
\centering
\includegraphics[width=\columnwidth]{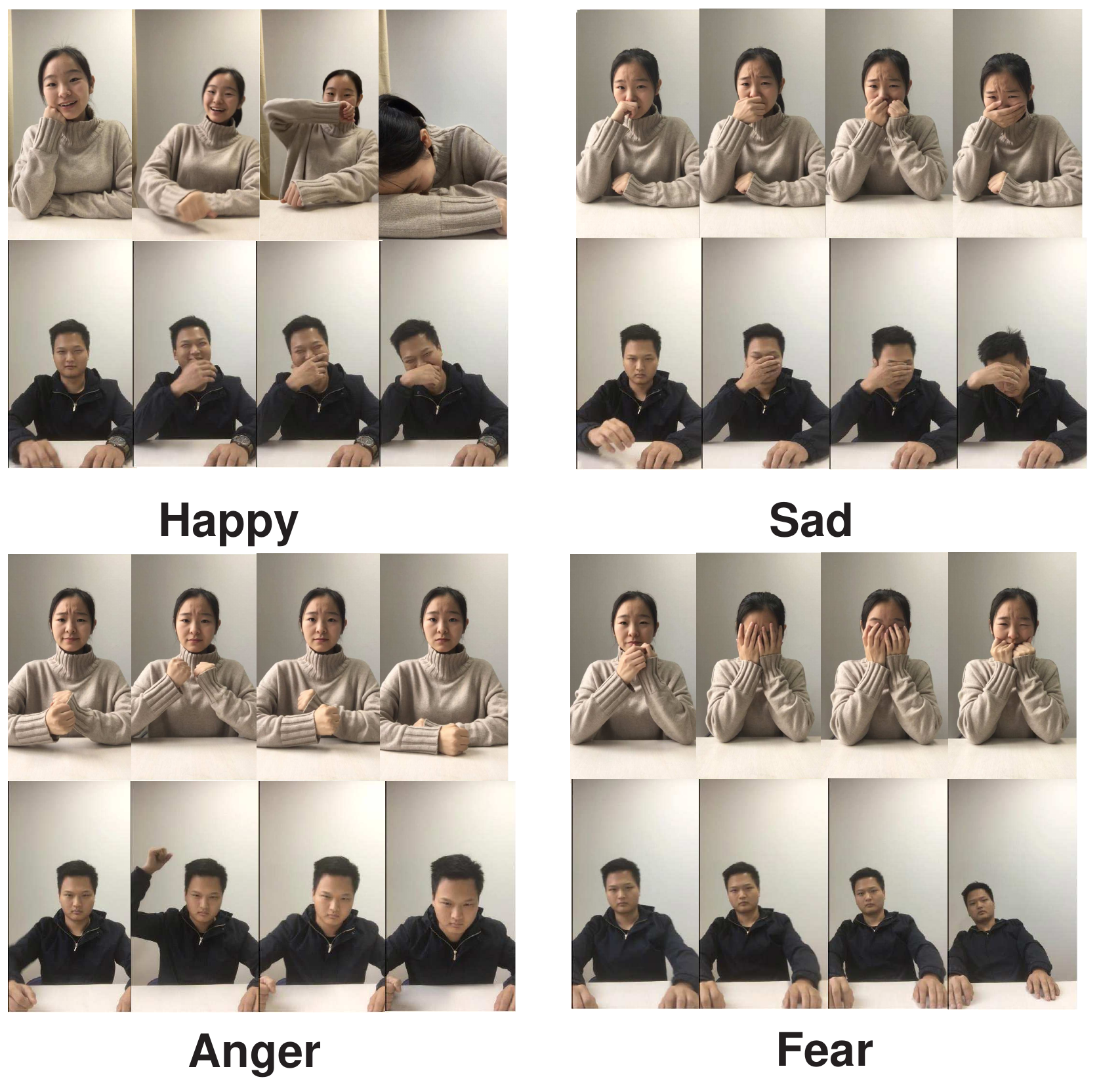}
\caption{The body gesture of emotion is person-dependent: an example.}
\label{fig:4Emotion}
\end{figure}
\noindent \textbf{\emph{3) The body gesture of emotion is person-dependent:}} The channel response data on physical expressions of emotions depend on persons examined. Fig. \ref{fig:EmotionIsPersonDependent} presents the CSI amplitude of three different participants watching the same clip of one-minute comedy. Even though all of them feel the same emotion (happiness), they have different expressions of it. For example, participant 3 is more dynamic with her data showing clear signal fluctuation in the bottom figure. Participant 3 is a female while the other two are males. The difference might be the result of different genders in expressing emotions, as females may be more expressive than males \cite{Zhao2016Emotion}.

{Fig. \ref{fig:4Emotion} shows such an example. Firstly, the expression of emotion is multi-modality via gestures, facial expression, and physiological signals. Secondly, the expression of emotion clearly depends on persons. But an interesting question is that whether such difference is related to genders, which will be studied in Section \ref{sect:PerEva}. }

In a word, through our pilot experiments, the relation between physical expressions of emotions and channel response data is confirmed, and meanwhile, two major challenges are to be responded to for designing EmoSense., i.e.

\begin{enumerate}
  \item How to adjust the experimental settings to enhance the fingerprint of the physical expressions of emotions?
  \item The fingerprint of channel response data on physical expressions of emotions is person-dependent. Then how can we distinguish the different emotions on different persons?
\end{enumerate}

In the following section, a Fresnel zone based look-up method will be proposed to respond to the first challenge as well as a data-driven architecture to take on the second one.

\section{System Design}
\label{sect:SysDe}
In this section, we first present a Fresnel zone based look-up method for adjusting the system setup to capture fine-grained gesture fingerprints on channel response. Then we offer a computational intelligence driven scheme for recovering the corresponding emotion from its body gestures.

\subsection{A  Fresnel Zone based Look-up Method}

Unlike previous similar research \cite{Sigg2014RF,Feng2016SleepSense,Zhao2016Emotion,Zheng2016Smokey,Wang2017RTfall} relying on the empirical experiences for system setup, we present  theoretic analysis  based on Fresnel Zones instead.

\begin{figure}
\centering
\subfigure[Fresnel Zone]{
\centering
\begin{minipage}[t]{0.4\textwidth}
\centering
\includegraphics[width=\columnwidth]{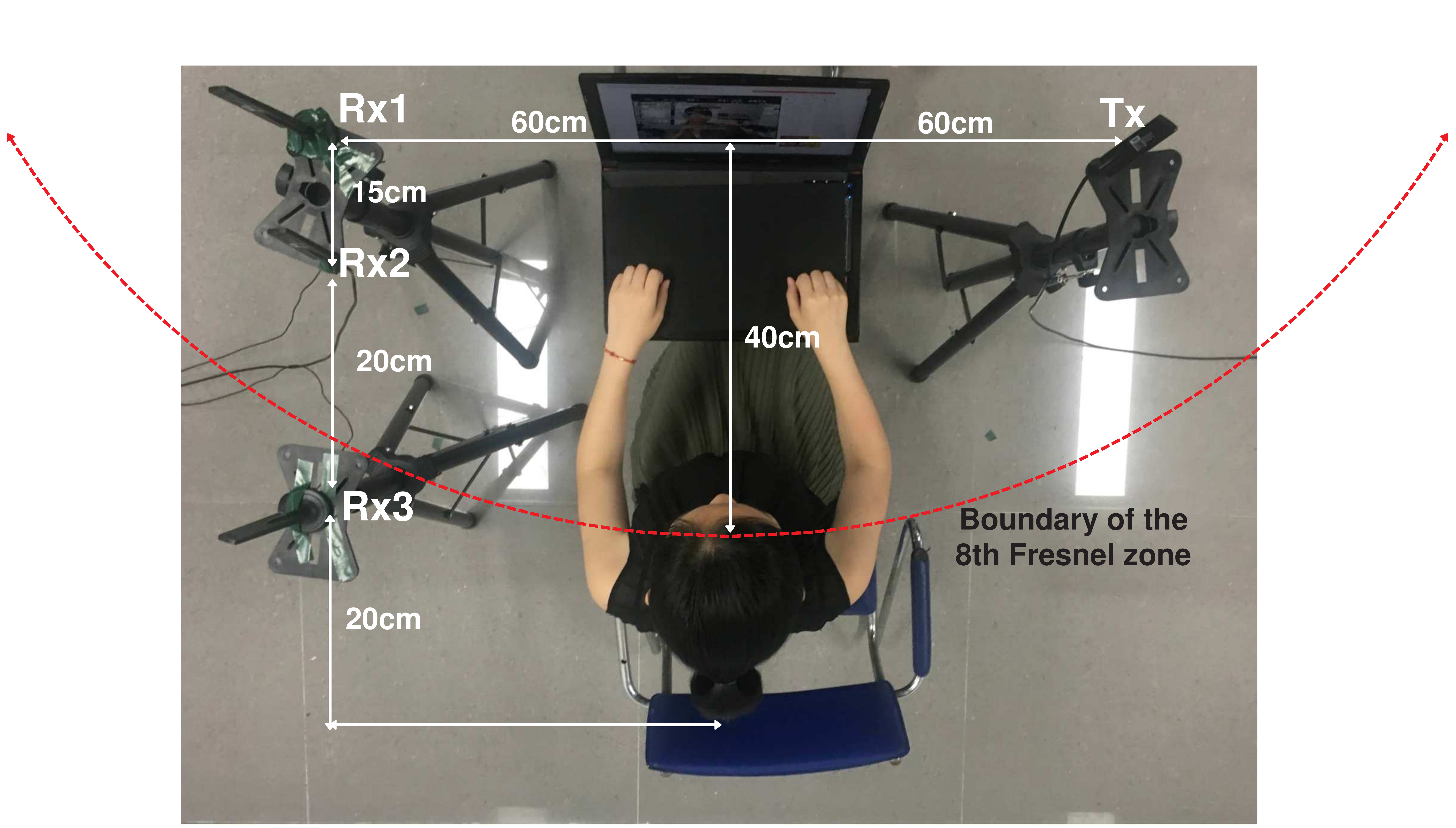}
\label{fig:FZoneTheory1}
\end{minipage}
}
\subfigure[Signal superposition]{
\centering
\begin{minipage}[t]{0.4\textwidth}
\centering
\includegraphics[width=\columnwidth]{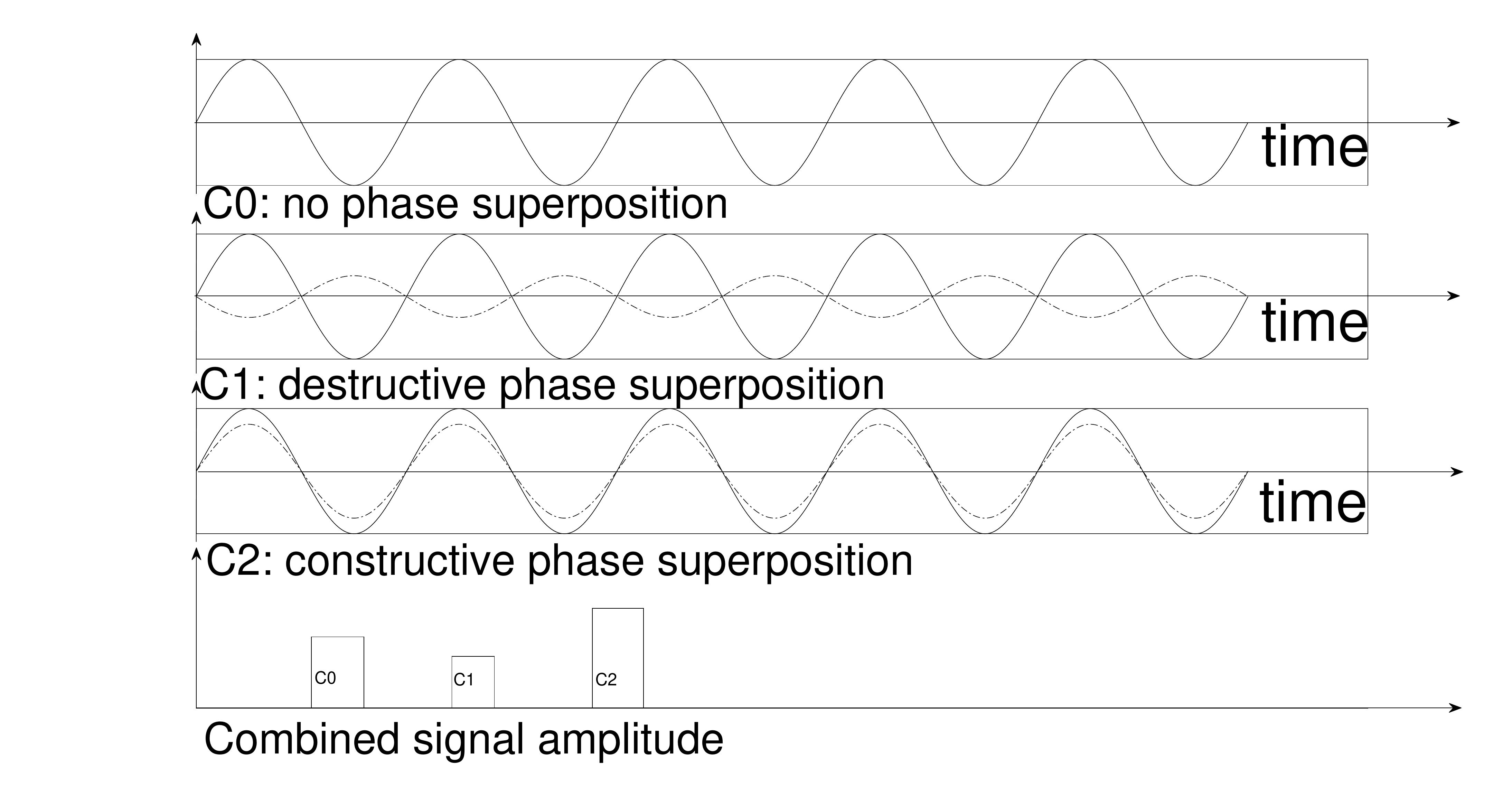}
\label{fig:FZoneTheory2}
\end{minipage}
}
\caption{An example of using Fresnel zones for enhancing the signal.}
\label{fig:FZone}
\end{figure}

Fig. \ref{fig:FZone} shows an example of the Fresnel zone, which consists of a set of concentric ellipsoids:

\begin{equation}
\frac{x^2}{a_n^2}+\frac{y^2}{b_n^2}  =1, n=1\cdots N\\
\end{equation}

\noindent where $Q_n(a_n,b_n)$ is a boundary point of the $n$th Fresnel zone.  $Tx$ and $Rx$ represent the sender and receiver, respectively.

For wireless signal with wavelength $\lambda$, the corresponding Fresnel zones can be constructed as follows,

\begin{equation}
|TxQ_n|+|Q_nRx|-|TxRx| = n \frac{\lambda}{2}
\label{eqn:4}
\end{equation}

WiFi signal, whether it is running on 2.4 GHz or 5 GHz, can hardly penetrate human beings. Therefore, a person acts like a mirror (reflector) to it, leading to a multi-path effect. In other words, signal collected at the receiver's end are from two types of paths: direct path (also named Line-of-sight path) and reflected path (also named as Non-line-of-sight) \cite{Zhang2017Centi}.

Let us particularly look at the phase shift $\Delta p$ of the received signal at $Rx$. Consider a person present at $Q_n$ on Fig. \ref{fig:FZone}, then to $Rx$ the LoS path is $Tx\rightarrow Rx$ while the NLoS path is $Tx\rightarrow Q_n \rightarrow Rx$. Clearly, the NLoS path is longer than the LoS path, and the difference is $|TxQ_n|+|Q_nRx|-|TxRx|= n \frac{\lambda}{2}$ according to Eqn. \ref{eqn:4}. This difference in distance induces a phase shift $\Delta p_1$ in signal:

\begin{equation}
\label{eqn:5}
\Delta p_1 =\left \{ \begin{array}{ll} 0, & \textrm{$n$ is even} \\
 \pi, & \textrm{$n$ is odd}
\end{array} \right.
\end{equation}

Moreover, a phase shift denoted as $\Delta p_1 = \pi$ is incurred when the signal is reflected. As a result, the combined phase shift $\Delta p$ at $Rx$  is $ \pi$ for the even Fresnel zone and $2\pi$ for the odd Fresnel zone:

\begin{equation}
\label{eqn:5}
\Delta p =\left \{ \begin{array}{ll} \pi, & \textrm{$n$ is even} \\
 2\pi, & \textrm{$n$ is odd}
\end{array} \right.
\end{equation}

Fig. \ref{fig:FZoneTheory2} demonstrates the combined signal. It is inspiring to see that the amplitude of the combined signal is degraded at the even zones and enhanced at the odd zones, during to the shifted phase.

The above observation urges us to leverage such phenomenon to envhance the impact of body gestures on channel response. The key idea is to adjust the layout of antennas to ensure that gesture happens in the odd Fresnel zones, so as to enhance its corresponding signal fingerprint.

To this end, we design a Fresnel zone based look-up method to guide the layout of antennas for better performance. If the subject locates at $Q_n$, the distance between $Q_n$ and $O$ can be calculated as follows,

\begin{equation}
\label{eqn:6}
\begin{array}{ll} |Q_nO|   = & \sqrt{|Q_nRx|^2-|ORx|^2} \\
                           = & \sqrt{(\frac{n\lambda}{2}+|TxRx|-|TxQ_n|)^2-|ORx|^2}, \\
                           = & \sqrt{(\frac{n\lambda}{4}+\frac{|TxRx|}{2})^2-|OTx|^2}, \\
                           = & \sqrt{(\frac{n\lambda}{4}+|OTx|)^2-|OTx|^2}, \\
                           = & \sqrt{\frac{n^2\lambda^2}{16}+\frac{n\lambda|TxRx|}{4}},

\end{array}
\end{equation}

\begin{table}
\centering
\caption{The look-up table}
\label{tab:LookUP}
\begin{tabular}{cc}
\hline
\multicolumn{1}{|c|}{n} & \multicolumn{1}{c|}{$|Q_nO|$}                                                                 \\ \hline
\multicolumn{1}{|c|}{0} & \multicolumn{1}{c|}{0}                                                                       \\ \hline
\multicolumn{1}{|c|}{1} & \multicolumn{1}{c|}{$\sqrt{\frac{\lambda^2}{16}+\frac{\lambda l}{4}}$} \\ \hline
\multicolumn{1}{|c|}{2} & \multicolumn{1}{c|}{$\sqrt{\frac{\lambda^2}{4}+\frac{\lambda l}{2}}$}                                                                        \\ \hline
\multicolumn{1}{|c|}{...} &\multicolumn{1}{c|} {...} \\ \hline
\multicolumn{1}{|c|}{n}   & \multicolumn{1}{c|}{$\sqrt{\frac{n^2\lambda^2}{16}+\frac{n\lambda l}{4}}$}  \\ \hline
\end{tabular}
\end{table}

The wavelength of WiFi signal under 2.4GHz and 5GHz is 2cm and 6cm, respectively. If we define the distance between $Tx$ and $Rx$ as $l$, a look-up table like Tab. \ref{tab:LookUP} can be constructed to set up the system quickly to ensure a better resolution.

\begin{figure}
\centering
\includegraphics[width=0.9\columnwidth]{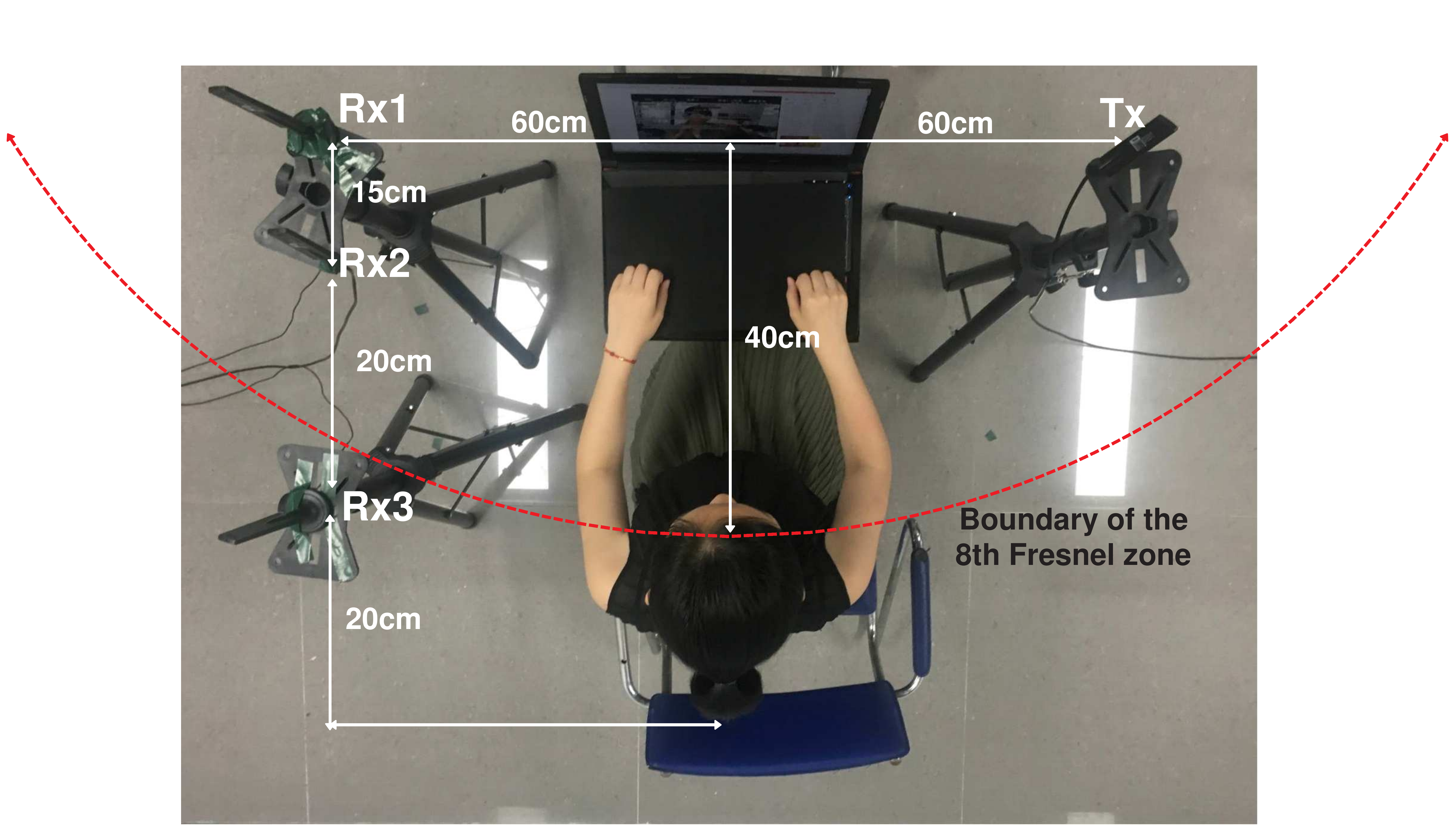}
\caption{The corresponding experimental setting.}
\label{fig:FZoneReal}
\end{figure}

Fig. \ref{fig:FZoneReal} shows one of our system setups as an example. The prototype consists of one transmitting antenna $Tx$ and three receiving antennas $Rx1$, $Rx2$ and $Rx3$. As the distance between $Tx$ and $Rx1$ is 120 cm, the look-up table points that the subject should be 40 cm away from this pair of transceiver, so that her gesture on channel response can be enhanced at the $8$th Fresnel zone \cite{Wang2016Human}. The locations of the rest two receiving antennas have been determined in a similar way.

\begin{figure}
\centering
\includegraphics[width=\columnwidth]{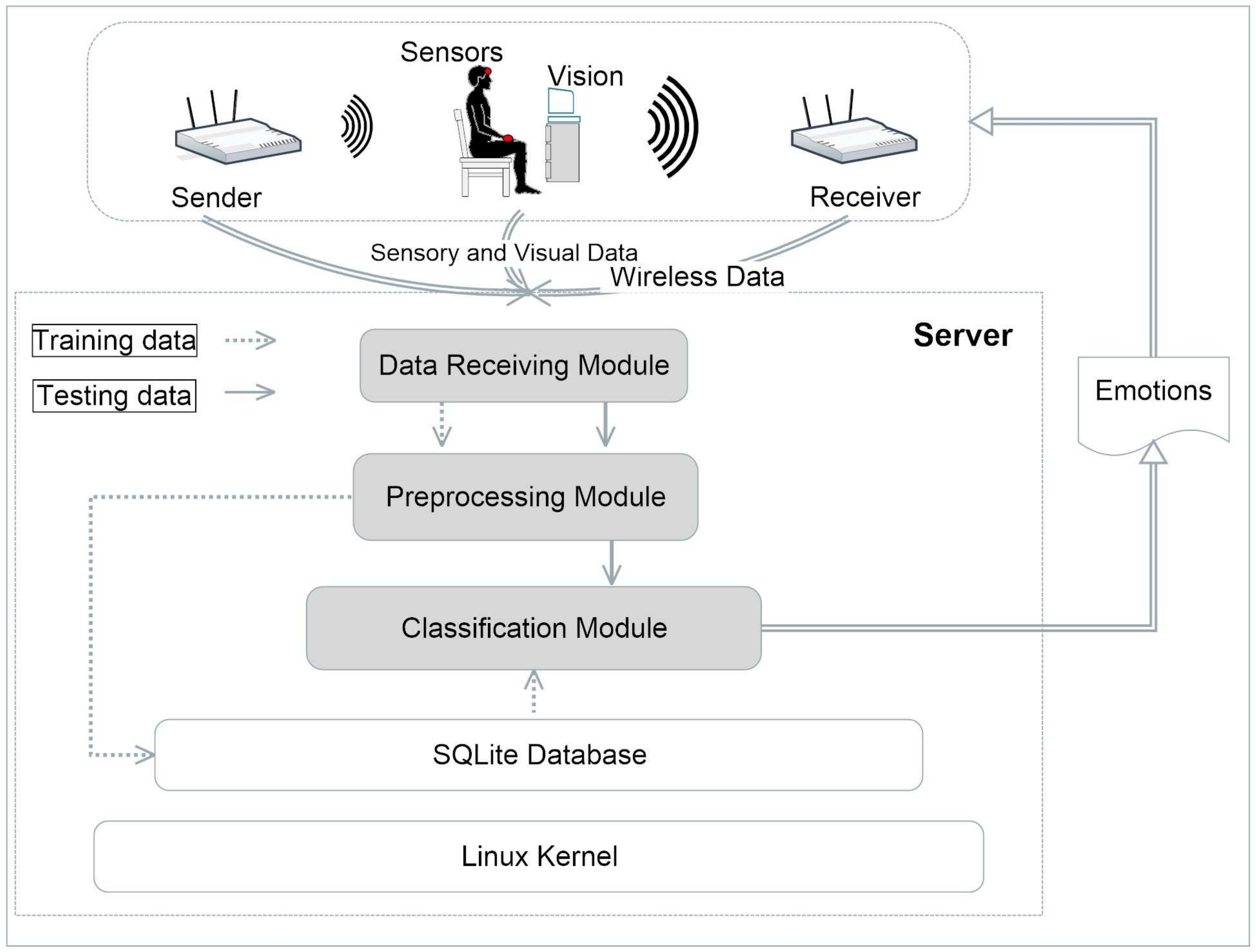}
\caption{System architecture of EmoSense.}
\label{fig:SysArt}
\end{figure}

\subsection{A Computational Intelligence driven Architecture}
\label{subsect:DataDriven}

Though the expression of emotions varies with persons, there still exist certain common patterns that can be explored, e.g., dancing for joy. This observation inspires us to develop a data-driven architecture leveraging computational intelligence to efficiently extract those patterns for emotion recognition, as shown in Fig. \ref{fig:SysArt}.

Fig. \ref{fig:SysArt} shows the system architecture of EmoSense. Like any typical data-mining system, EmoSense also relies on mining the data for emotion recognition. Therefore, the training data is essential. It flows from the data receiving module to the preprocessing module for interpolation, denoising and feature extraction, and then reaches the SQLite database. After the training phase, EmoSense is online for testing. The testing data also originates from the data receiving module to the preprocessing module, and then reaches the classification module for emotion recognition.

\noindent [\textbf{Preprocessing Module}]. The raw data may be not complete due to information loss on the noisy channel. Therefore, we first correct this issue via a commonly-used linear interpolation technique. Then we filter the raw data with a Butterwoth filter \cite{Ali:2015:WiKey}. The cut-off frequency $\omega_c$ of the Butterworth filter is set to $\omega_c = \frac{2\pi\cdot f}{F_s}=\frac{2\pi\cdot 15}{100}=0.942$ rad/s, where $F_s$  represents the sampling rate (100 samples per second in our system).

\noindent [\textbf{Classification Module}]. This module leverages temporal-frequency features extracted from the gesture fingerprint to deduct the corresponding emotion. Here three classic classifiers, i.e.,  k-NN, NaiveBayes, and Bagging, are used.

\section{Performance Evaluation}
\label{sect:PerEva}

In this section, we will conduct an exhaustive evaluation of the performance of EmoSence.

\subsection{Evaluation Setup}

A prototype system of EmoSense is built and evaluated in the real environment setting, which is a   $7\times 10 m^2$ office, as shown in Fig. \ref{fig:prototype}.  It has office furniture, such as couches, chairs, computer tables and book shelves. Some students are also in the same office doing their works during the experiments, for the purpose of providing a real-world environment.

\noindent \textbf{[Metric].}  A confusion matrix containing the overall accuracy (cf. \cite{Sigg2014RF,Gu16IoT,Gu17IoT,Zhao2016Emotion}) is used to evaluate the overall performance of EmoSense.

\noindent \textbf{[Data Set].} For each emotion, we define it from three difference motion sequences. And each participant is required to perform 20 times of each sequence. Then, we have  $14\times 3\times 20=3360$ data entries for the data set.

\noindent \textbf{[Feature].} As in \cite{samanta2003artificial}, seven features in both time and frequency domains are been selected, namely,

\begin{itemize}
  \item Standard deviation.  $\rho = \sqrt{\frac{1}{N}\sum_{i=1}^N (x_i-\mu)^2}$
  \item Average absolute error. $\Delta=\frac{1}{N}\sum_{i=1}^N|\Delta_i|$
  \item Skewness. $Skew(X)=E[(\frac{X-\mu}{\rho})^3]$
  \item Kurtosis. $K=\frac{\sum_{i=1}{K}(x_i-x)^4f_i}{ns^4}$
  \item Entropy. $H(X)=-\sum_i P(x_i)\log_b P(x_i)$
  \item Standard deviation of the velocity of the signal changing.
  \item Median. $M(X)= \left \{ \begin{array}{ll} x_{\frac{n+1}{2}}, & \textrm{$n$ is odd} \\
\frac{x_\frac{n}{2} + x_{\{\frac{n}{2}+1\}}}{2}, & \textrm{$n$ is even}
\end{array}\right.$
\end{itemize}

\noindent \textbf{[Classifier].} Three classic classifiers have been used, i.e., k-Nearest Neighbor (k-NN), Support Vector Machine (SVM) and Naive Bayes.

\subsection{Main-stream Benchmarks}
We design a sensor-based and a vision-based system to capture the same physical expression as EmoSense does as performance benchmarks.

\begin{figure}
\centering
\subfigure[The sensor-Based system]{
\includegraphics[width=40mm]{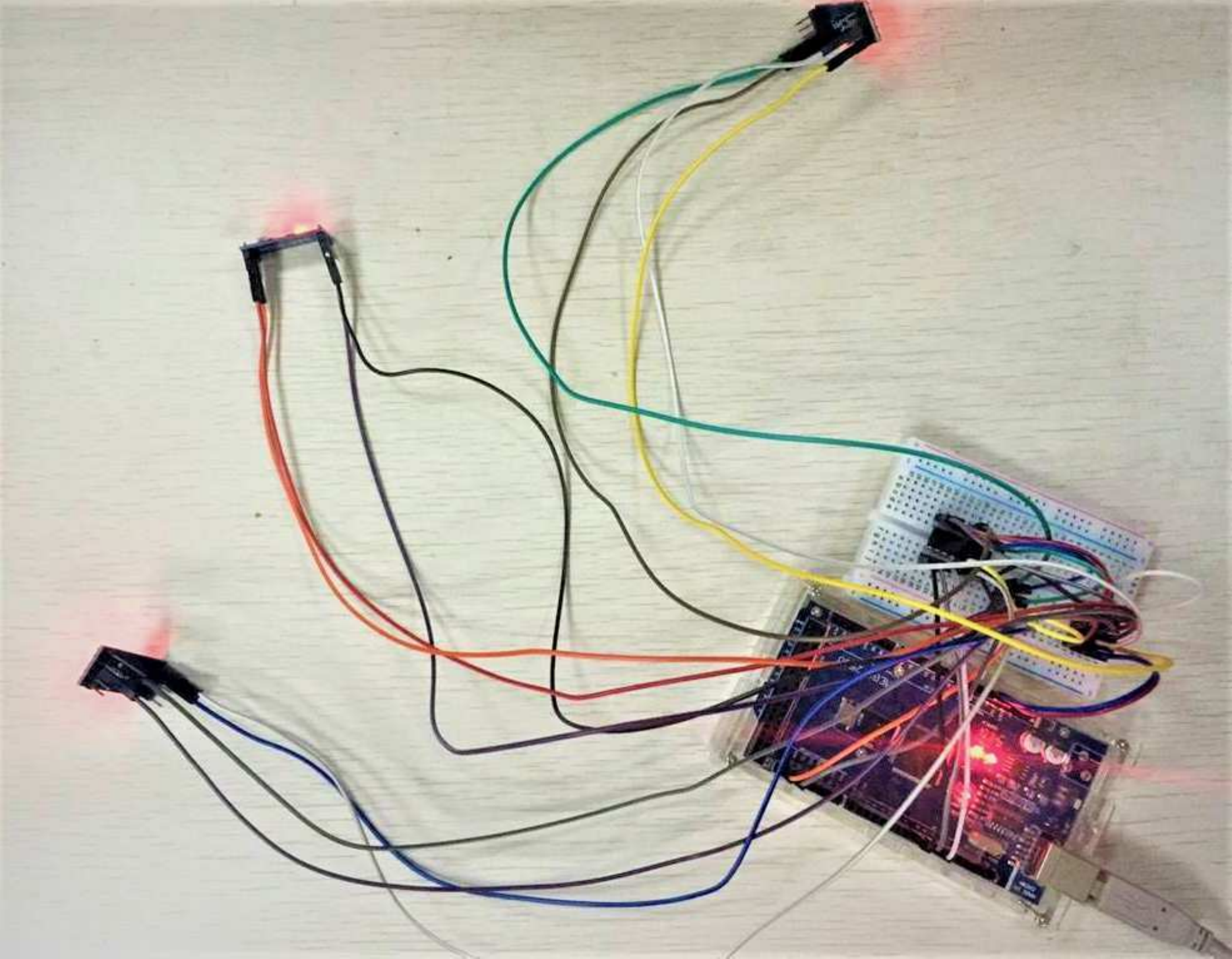}
\label{fig:arduino}
}
\subfigure[The vision-based system]{
\includegraphics[width=40mm]{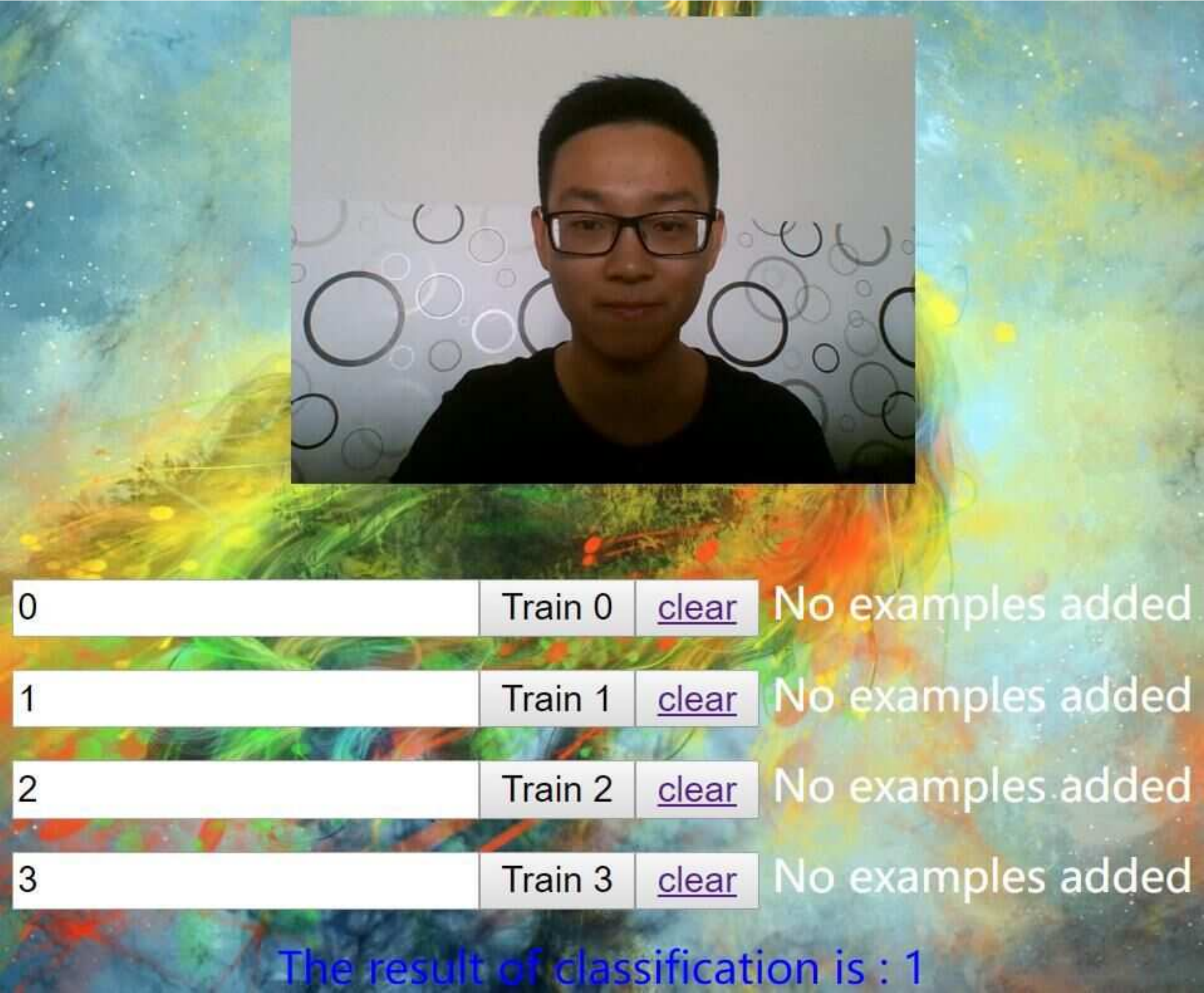}
\label{fig:vision}
}
\caption{Two Mainstream Benchmarks.}
\label{fig:benchmark}
\end{figure}

\noindent \textbf{[Sensor-based].} The sensor-based system is built upon an Arduino platform mounted with three ADXL345 accelerometer sensors shown in Fig. \ref{fig:arduino}. They are attached to the forehead, and two wrists of the subject, respectively. We use the same features and classifiers for both EmoSense and the sensor-based system.

\noindent \textbf{[Vision-based].} The vision-based system is based on an open-source machine learning library `deeplear.js' released by Google as shown in Fig. \ref{fig:vision}. The built-in camera of the laptop is used to capture both the facial and physical expression of the subject.  Each frame will first be processed by a mini neural network named SqueezeNet, and the penultimate layer in the neural networks will be utilized for training and testing with a KNN classifier.

\subsection{Overall Evaluation}
In this part, we present and analyze the evaluation results, which are achieved through ten-fold cross-validation.

\begin{table}
\centering
\caption{{Confusion Matrix of k-NN algorithm: Inset Test (Upperbound)}}
\label{tab:Inset}
\resizebox{90mm}{22mm}{
\begin{tabular}{|c|c|c|c|c|c|}
\hline
\multicolumn{2}{|c|}{ } & HAPPY(\%)   & SAD(\%)     & ANGER(\%)   & FEAR(\%)    \\

\hline
\multirow{3}*{HAPPY(\%)} &KNN                   & \textbf{100}& 0 & 0 & 0  		\\  \cline{2-6}  
			               &SVM                   & \textbf{85.71}& 7.14& 2.38& 4.77  		\\  \cline{2-6}  
                                       &NaiveBayes       & \textbf{32.38}& 6.67& 44.29& 16.66  		\\  \cline{2-6}  
\hline

\multirow{3}*{SAD(\%)}      &KNN                  & 2.86&\textbf{97.14}& 0&0  		\\  \cline{2-6}  
			              &SVM                  & 10.48& \textbf{74.29}&3.33& 11.90 		 \\  \cline{2-6}  
                                      &NaiveBayes       & 21.43&\textbf{19.52}&48.57& 10.48 		 \\  \cline{2-6}  
\hline

\multirow{3}*{ANGER(\%)} &KNN                 &1.90&5.71&\textbf{ 92.39}&0 		\\  \cline{2-6}  
			               &SVM                 & 12.86& 9.05&\textbf{70.48}&7.61 		 \\  \cline{2-6}  
                                       &NaiveBayes      &15.24&7.14& \textbf{70.95}&6.67		 \\  \cline{2-6}  
\hline
\multirow{3}*{FEAR}    &KNN                 &6.67&7.62&1.90&\textbf{83.81}  		\\  \cline{2-6}  
			               &SVM                 & 9.52&10.00&2.38&\textbf{78.10} 		 \\  \cline{2-6}  
                                      &NaiveBayes     &24.76&4.76&47.14&\textbf{23.34} 		 \\  \cline{2-6}  
\hline
\multicolumn{5}{|r|}{KNN Avg.}  & 93.33\% \\ \hline
\end{tabular}
}
\end{table}

\begin{table}
\centering
\caption{{Confusion Matrix of k-NN algorithm: Person-dependent}}
\label{tab:PerDep}
\resizebox{90mm}{22mm}{
\begin{tabular}{|c|c|c|c|c|c|}
\hline
\multicolumn{2}{|c|}{ } & HAPPY(\%)   & SAD(\%)     & ANGER(\%)   & FEAR(\%)    \\

\hline
\multirow{3}*{HAPPY(\%)} &KNN                   & \textbf{84.29}&5.24&6.19&4.28		\\  \cline{2-6}  
			               &SVM                   & \textbf{80.59}&9.52&2.38&7.15  		\\  \cline{2-6}  
                                       &NaiveBayes      & \textbf{30.95}&8.57&44.29&16.19  		\\  \cline{2-6}  
\hline

\multirow{3}*{SAD(\%)}      &KNN                   & 3.18&\textbf{83.33}&3.33&9.53 		\\  \cline{2-6}  
			              &SVM                    &12.86&\textbf{69.05}&3.81&14.28 		 \\  \cline{2-6}  
                                      &NaiveBayes       &22.38&\textbf{19.05}&47.62&10.95		 \\  \cline{2-6}  
\hline

\multirow{3}*{ANGER(\%)} &KNN                 &2.38&5.24&\textbf{89.05}&3.33		\\  \cline{2-6}  
			               &SVM                  &13.81&10.48&\textbf{68.57}&7.14 		 \\  \cline{2-6}  
                                       &NaiveBayes     &15.24&8.10& \textbf{70.48}&6.18 		 \\  \cline{2-6}  
\hline
\multirow{3}*{FEAR(\%)}    &KNN                 &7.14&8.10&1.90&\textbf{82.86}  		\\  \cline{2-6}  
			               &SVM                 &13.33&10.48&2.38&\textbf{73.81} 		 \\  \cline{2-6}  
                                      &NaiveBayes     &23.81\%&7.14&46.19&\textbf{22.86} 		 \\  \cline{2-6}  
\hline
\multicolumn{5}{|r|}{KNN Avg.}  &84.88\% \\ \hline
\end{tabular}
}
\end{table}

\noindent \textbf{[Inset Classification].} The inset testing is to use all the data sets for training. Its result serves as the upperbound of the performance. Table \ref{tab:Inset} shows the corresponding results, where EmoSense with kNN has the best performance : $93.33$\% ACC, while the performance degeneration of SVM  over kNN on the same dataset is about 15\% to 20\%. It is quite interesting that Naive Bayes is much worse than its two rivals by achieving only \% ACC. We think the reason lies in two folders: 1) The independence assumption of Naive Bayes is not quite suitable for our case; 2) Naive Bayes is particularly sensitive to the initial training dataset.

\begin{figure}
\centering
\subfigure[Happy Performance in Signal]{
\includegraphics[width=42mm]{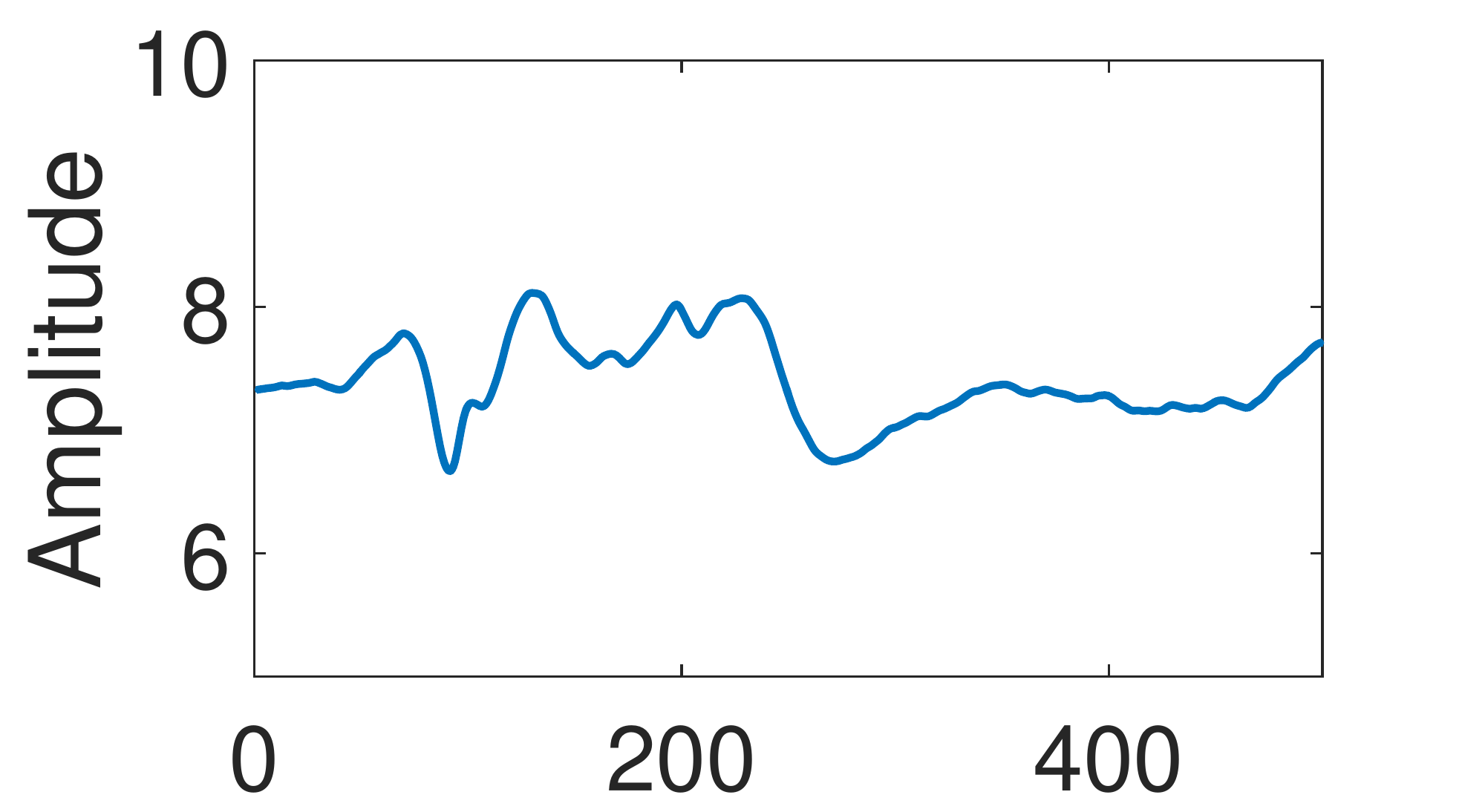}
\label{fig:1}
}
\subfigure[Sad Performance in Signal]{
\includegraphics[width=40mm]{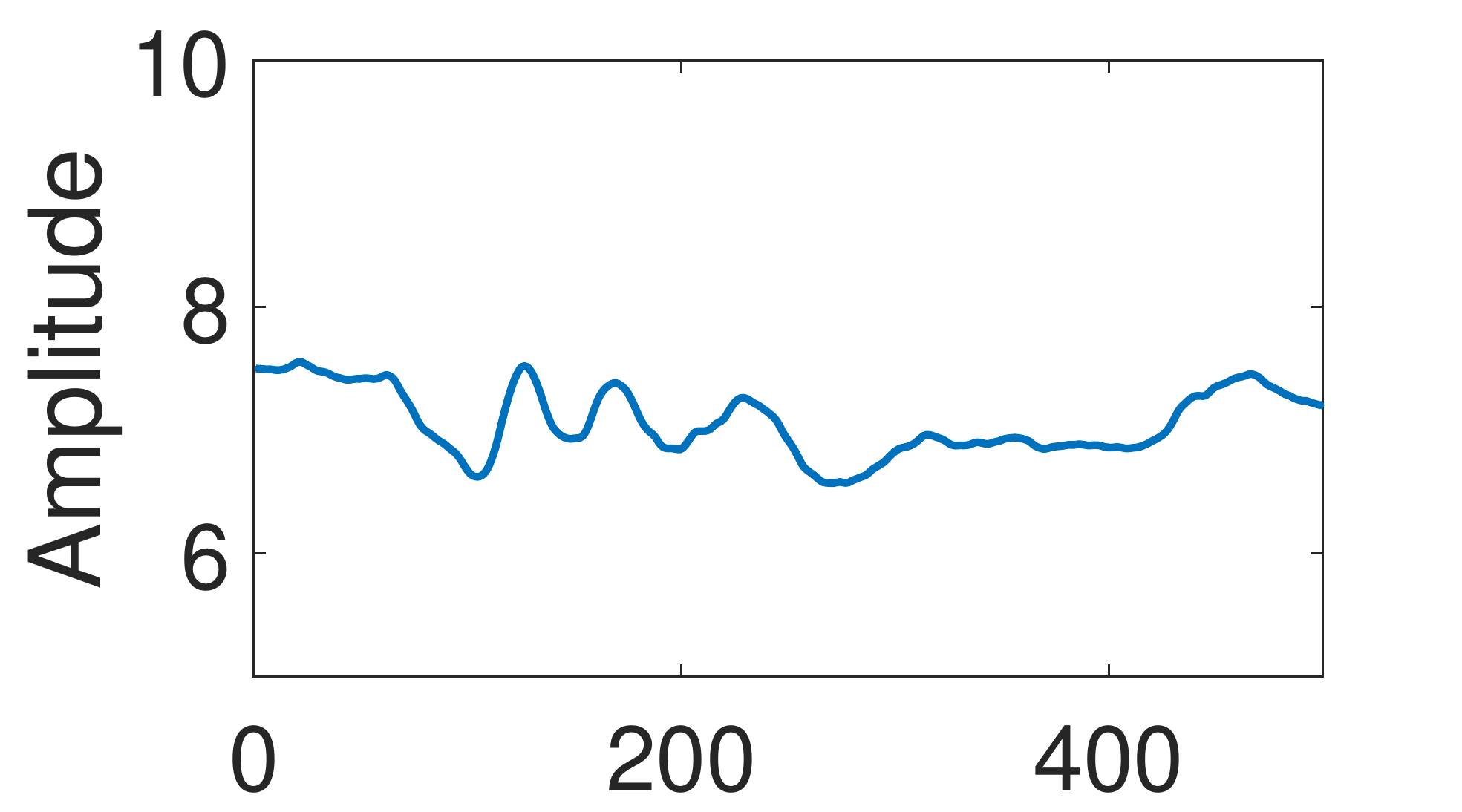}
\label{fig:2}
}
\subfigure[Anger Performance in Signal]{
\includegraphics[width=40mm]{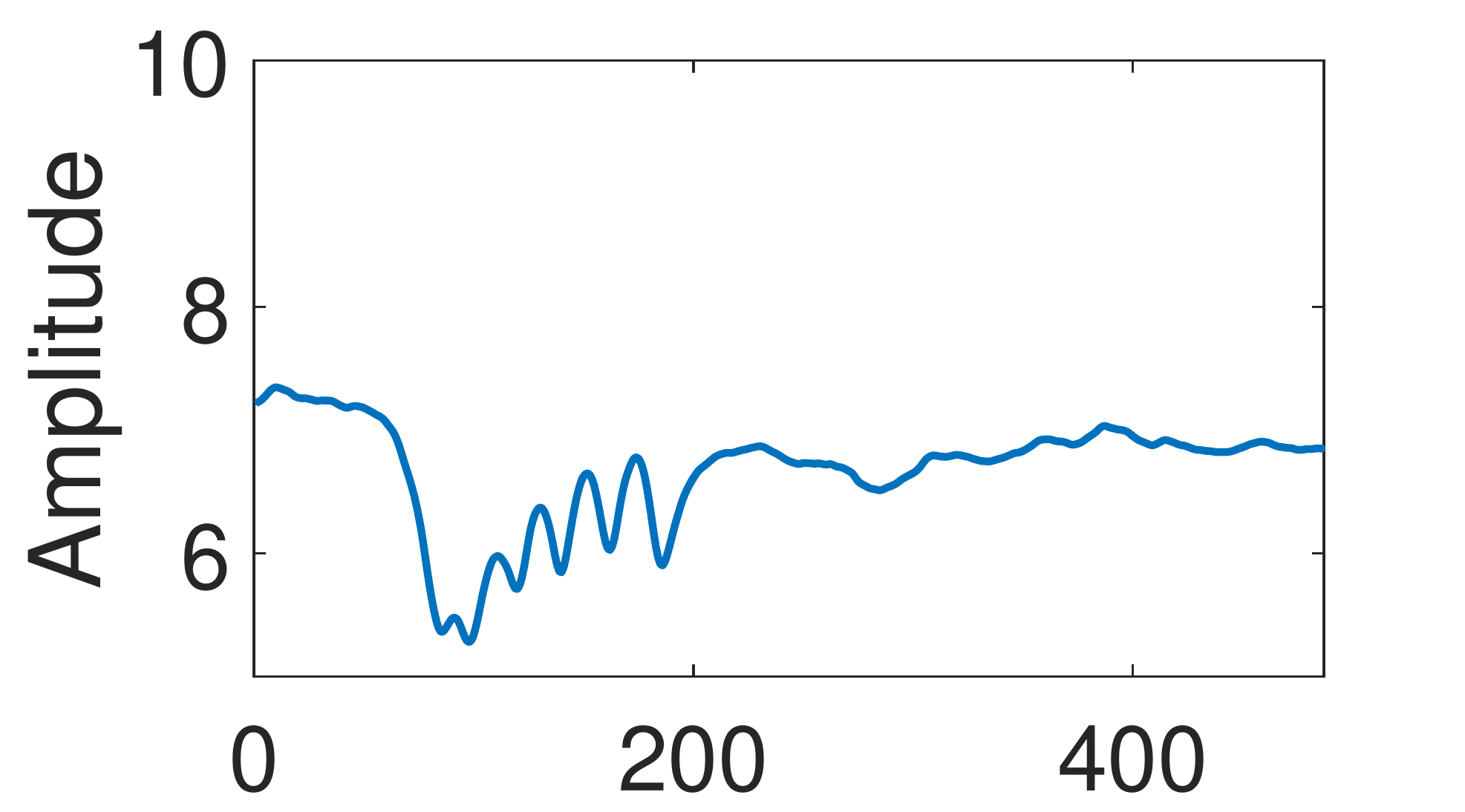}
\label{fig:3}
}
\subfigure[Fear Performance in Signal]{
\includegraphics[width=40mm]{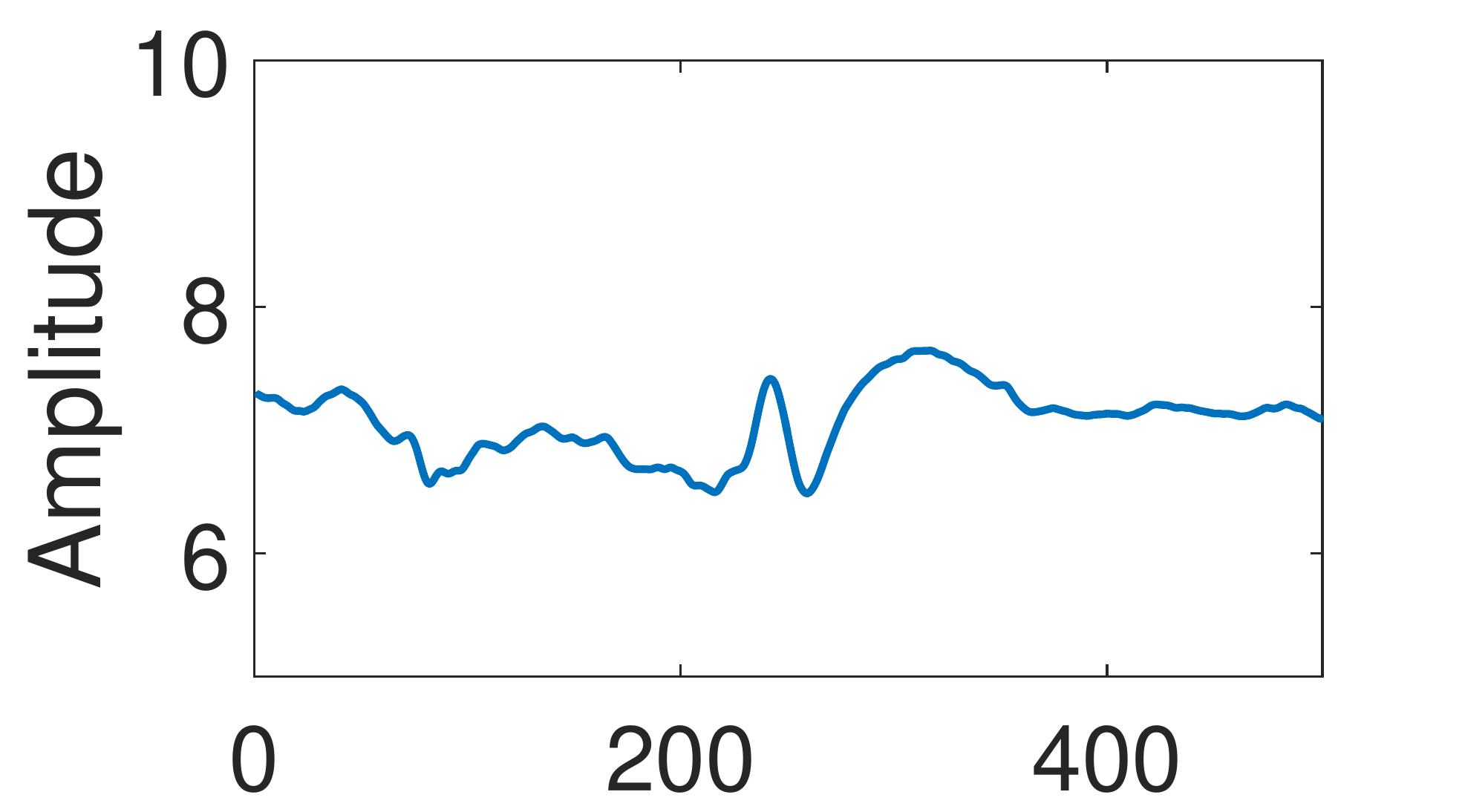}
\label{fig:4}
}
\caption{Emotions Performance Compare.}
\label{fig:compare in signal}
\end{figure}

Specifically, happy has the highest recognition ratio (100.00\%)  while fear is the hardest to identify (83.81\%). In other words, happy has never been misinterpreted as other emotions, and it is the most distinguishable. The video footage has been carefully examined to understand such phenomenon. It is found that fear has the least intensive expression while happy has the most. It is also interesting that fear sometimes can be misjudges as happy (6.67\%). For instance, Fig. \ref{fig:compare in signal} shows different emotions in signal for the same subject. We can see happy and anger have larger signal fluctuation (intensity) compared to sad and fear, which means that they contain more significant physical expressions.

\noindent \textbf{[Person-dependent Classification].} In this case, we select partial data of some subject for training and use the rest for testing. The results are concluded in in Table \ref{tab:PerDep}, where EmoSense achieves 84.88\% ACC, which is close to the performance upper bound (93.33\%). k-NN is still the best among three classifiers. Fear still has the worst recognition accuracy among all four emotions. But anger has the best performance (89.05\%) here rather than happy as in the in-set classification. The phenomenon is consistent with Fig. \ref{fig:compare in signal}. The results confirm that there exist certain common patterns of expression for emotion across different persons.

\begin{table}
\centering
\caption{{Confusion Matrix of k-NN algorithm: Person-independent}}
\label{tab:PerIndep}
\resizebox{90mm}{22mm}{
\begin{tabular}{|c|c|c|c|c|c|}
\hline
\multicolumn{2}{|c|}{ } & HAPPY(\%)   & SAD(\%)     & ANGER(\%)   & FEAR(\%)    \\

\hline
\multirow{3}*{HAPPY(\%)} &KNN                   & \textbf{46.43}&16.31&12.62&24.64		\\  \cline{2-6}  
			               &SVM                   & \textbf{52.74}&16.55&15.11&15.60  		\\  \cline{2-6}  
                                       &NaiveBayes      & \textbf{22.14}&5.83&48.45&23.58  		\\  \cline{2-6}  
\hline

\multirow{3}*{SAD(\%)}      &KNN                   &14.76&\textbf{35.48}&17.02&32.74\ 		\\  \cline{2-6}  
			              &SVM                    &22.38&\textbf{41.67}&10.83&25.12 		 \\  \cline{2-6}  
                                      &NaiveBayes       &21.19&\textbf{11.07}&52.14&15.60 		 \\  \cline{2-6}  
\hline

\multirow{3}*{ANGER(\%)} &KNN                 &21.19&26.07&\textbf{41.07}&11.67		\\  \cline{2-6}  
			               &SVM                  &24.64&28.21&\textbf{38.81}&8.34 		 \\  \cline{2-6}  
                                       &NaiveBayes     &18.21&13.10& \textbf{56.07}&12.62 		 \\  \cline{2-6}  
\hline
\multirow{3}*{FEAR(\%)}    &KNN                 &30.71&22.02&7.50&\textbf{39.77}  		\\  \cline{2-6}  
			               &SVM                 &33.33&23.57&4.17&\textbf{38.93} 		 \\  \cline{2-6}  
                                      &NaiveBayes     &27.02&4.52&47.86&\textbf{20.60} 		 \\  \cline{2-6}  
\hline
\multicolumn{5}{|r|}{KNN Avg.}  &40.86\% \\ \hline
\end{tabular}
}
\end{table}

\noindent \textbf{[Person-Independent Classification].} For person-independent classification, we exclude one subject's data set from the training set and use it for testing. The result is the average of all experimenters. As expected, the performance of EmoSense degenerates significantly. As shown in Table \ref{tab:PerIndep}, EmoSense achieves only $40.86$\% accuracy on average. We find that happy still is the highest (46.43\%) while sad becomes the lowest(35.48\%).  The huge performance deterioration of all four emotions  over the previous case implies that the expression of emotion indeed in person-dependent. In particular, sad has the largest performance degeneration, indicating that its expression heavily relies on the subject.

\begin{figure}[H]
\centering
\includegraphics[width=\columnwidth]{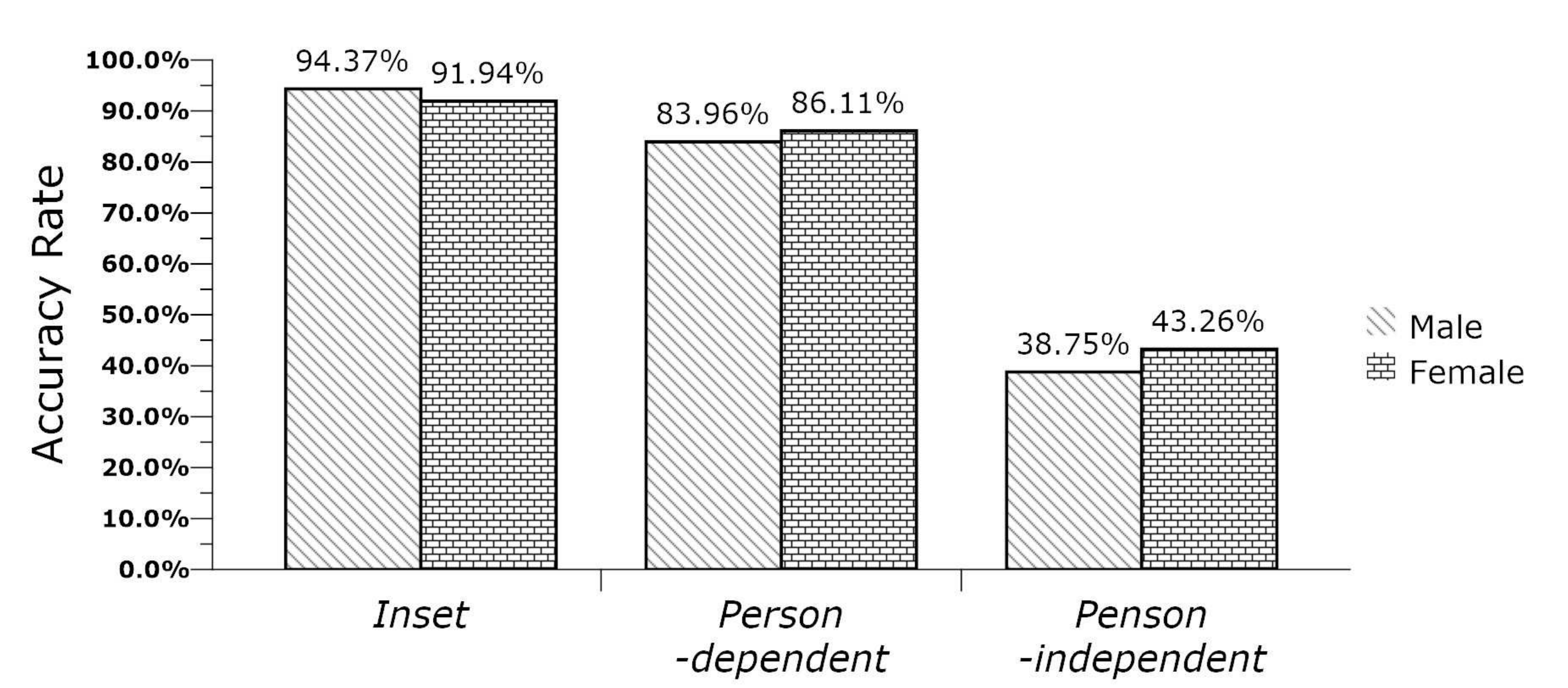}
\caption{Gender-based Classification.}
\label{fig:gender}
\end{figure}

\noindent \textbf{[Gender-based Classification].} Fig. \ref{fig:gender} shows the impact of gender on the overall system performance. For the inset test, both genders achieve high averaged accuracy, i.e., 94.37\% and 91.94\%, while male is slightly better. For the person-dependent and person-independent cases, female slightly outperforms male, i.e, 86.11\% vs 83.96\% and 43.26\% vs 38.75\%. In other words, we have not observed any significant differences between genders as suggested in other references \cite{Zhao2016Emotion}. {One possible reason is that the studied emotions are normal in our daily lives, where both genders share certain similarities in the expression. In the future work, we will involve more emotions and participants to further clarify this issue.}

\begin{figure}
\centering
\includegraphics[width=\columnwidth]{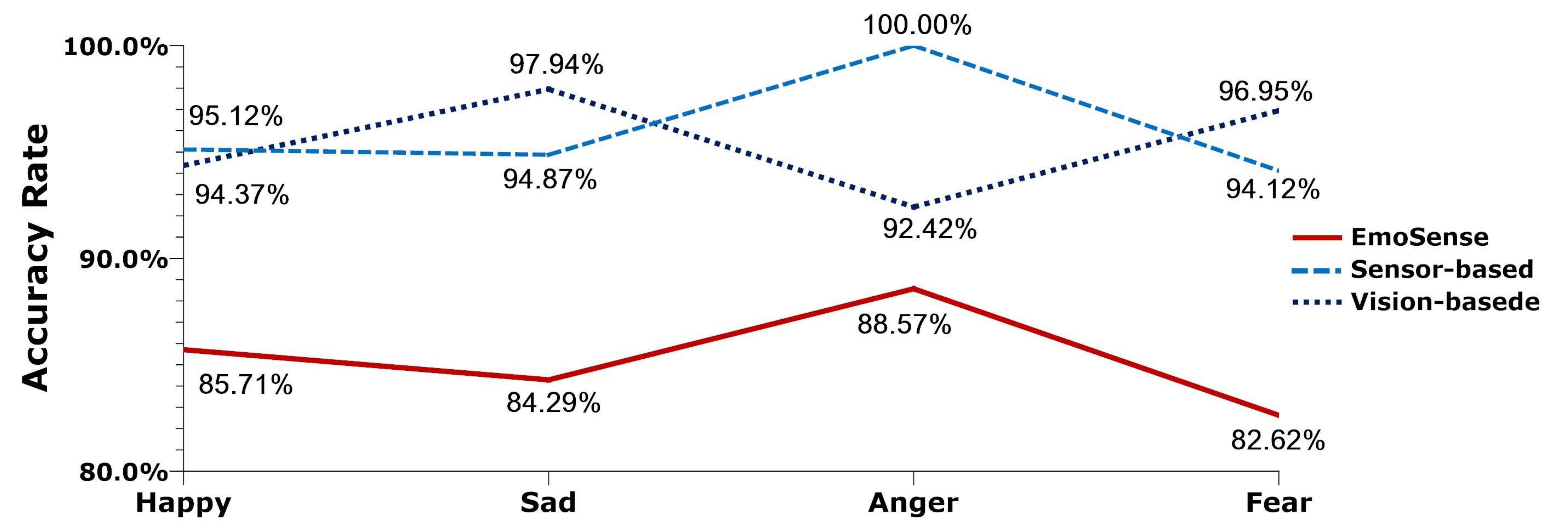}
\caption{Emosense and Benchmarks.}
\label{fig:benchmark}
\end{figure}

\subsection{Evaluation via State-of-the-art}
\noindent \textbf{[EmoSense versus Sensor-based system].} Sensors are usually contact and thus posses much less noisy than the wireless signal. Therefore, they are considered as a more reliable data source for recognition. It should serve as a golden performance for EmoSense. Since both systems record body gesture of emotion, we use the same classifiers and features for a fair comparison.

As expected, the sensor-based solution outperforms EmoSense in all four emotions by achieving 95.12\%, 94.87\% , 100\% and 94.12\% accuracy, respectively. It can be observed very intuitively that  the two fold lines of sensor based and EmoSense have the same transformation trend, and the performance gap in each emotion are all  maintain only at around 10\%.  It suggests that our idea of exploring the physical expression for emotion sensing is proper and the gap is mainly caused by the background noise of wireless signal. The results also verify our previous observation that fear has the worst performance, 91.12\%, while  happy and anger are better, 92.12\% and 100\%.

\noindent \textbf{[EmoSense versus Vision-based system].} Fig. \ref{fig:benchmark} indicates that the vision-based solution performs better than EmoSense, and the performance gap is about 10\%. But unlike in Emosense where sad can be hardly recognized, here it has the best performance (97.94\%) and the performance gap between two systems is 13.65\%. Because the expression of sad usually concentrates on the face than the body. Therefore, the vision-based system leveraging the facial expression can recognize sad accurately.

\emph{\textbf{Summary}}. Our early study \cite{Gu2018ICC} has already confirmed the feasibility of leveraging wireless signal for gesture and emotion recognition. Here, we not only confirmed its feasibility again, but also verified the proposed EmoSense system via extensive comparative experiments. As a result, we belive that EmoSense, which provides a reliable and transparent fall sensing service, constitutes attempting emotion sensing solution in real-world.

\begin{figure}
\centering
\includegraphics[width=\columnwidth]{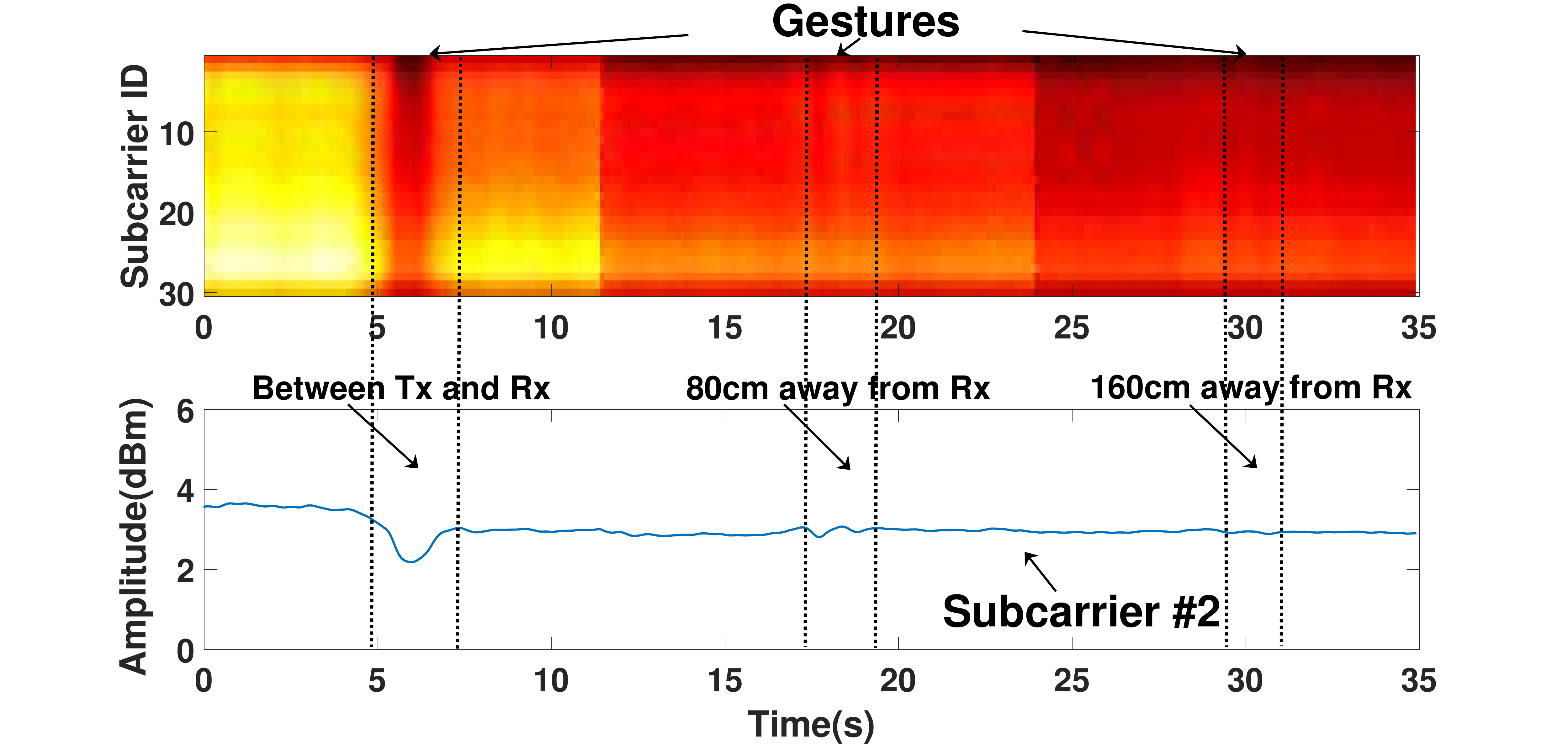}
\caption{A case study showing the robustness of EmoSense to the noises}
\label{fig:11}
\end{figure}

\subsection{Further Discussions}
{\noindent\textbf{[Ambient noise]}.  The ambient noise caused by nearby devices and persons could affect the performance of EmoSense. In our previous work} \cite{Gu17IoT}, {we have already shown that the interference from other wireless devices is quite limited. Here, we study the robustness issue to ambient noises caused by nearby humans.  To this end, a new experiment has been conducted as follows:}

    \begin{itemize}
      \item {One male participant performs the same hand gesture (waving up and down) at three different locations: between the \emph{Tx} and \emph{Rx}, $80$ cm away from \emph{Rx}, and $1.6$ m away from \emph{Rx}.}
    \end{itemize}

{Fig. \ref{fig:11} records the CSI amplitude data. As pointed out in the figure, the same hand gesture has totally different impacts on channel data at different locations. The closer to $Tx$ and $Rx$, the better the gesture can be captured. The gesture performed at the third location can hardly affect channel response. In other words, EmoSense is robust to noises caused by surrounding persons.}

{\noindent\textbf{[Real-world Application]}. Currently, EmoSense is limited in practice since
only four emotions can be recognized. But its system architecture is essentially data-driven, which could be extended for more emotions that do have physical expressions. This is one of key directions for the future work.}

{Even for the current EmoSense system, there may exist some real-world applications. For instance, there will be rehearsals before the first stage performance of a comedy. The club will set up the price of the ticket based on the reaction of the audience. EmoSense can be used for the exact purpose without any privacy concerns.}

\section{Conclusion and Future Work}
\label{sect:Conclusion}
In this paper, we present EmoSense, a first-of-its-kind wireless emotion sensing system driven by computational intelligence. It has been  prototyped on off-the-shelf WiFi devices and evaluated in real environments. Two traditional rivals, i.e., visions-based and sensor-based, have been realized for the comparative study. Performance evaluation over 3360 cases suggests that EmoSense achieves a comparable performance to the vision-based and sensor-based rivals under different scenarios with a classic k-Nearest Neighbor (kNN) classifier.

For the future work, there exists several open issues. Firstly, EmoSense and its rivals hinge upon human gestures as expression of emotion, which still remains a blur by far. {For example, dishonest people can deceive the system by intentionally behaving in certain ways. A possible solution is to leverage the multi-modality feature of emotion.} Secondly, EmoSense is data-driven. But it is a common sense that psychology knowledge is also very important. Therefore, it is more reasonable to couple both data and psychology knowledge for more reliable and accurate emotion recognition. Last but not least, the physical expression of emotion is affected by many congenital and acquired factors, some of which are totally out of control. Therefore, it is important to clarify the potential scenarios before we actually deploy the system.

\section*{Acknowledgments}
This work is sponsored by the National Natural Science Foundation of China (NSFC) under Grant No. 61772169, National Key Research and Development Program under Grant No.2018YFB0803403, the Fundamental Research Funds for the Central Universities under No.JZ2018HGPA0272, and  Open Projects by Jiangsu Province Key Laboratory of Internet of Things under No.JSWLW-2017-002.

\bibliographystyle{IEEEtran}
\bibliography{LoMD}

\begin{IEEEbiography}[{\includegraphics[width=1in,height=1.25in,clip,keepaspectratio]{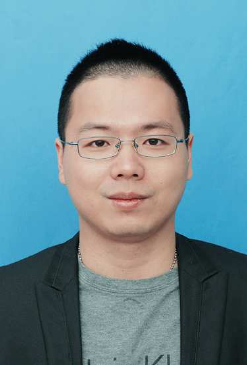}}]{Yu Gu} (M'10-SM'12)received the B.E. degree from the Special Classes for the Gifted Young, University of Science and Technology of China, Hefei, China, in 2004, and the D.E. degree from the same university in 2010.In 2006, he was an Intern with Microsoft Research Asia, Beijing, China, for seven months. From 2007 to 2008, he was a Visiting Scholar with the University of Tsukuba, Tsukuba, Japan. From 2010 to 2012, he was a JSPS Research Fellow with the National Institute of Informatics, Tokyo, Japan. He is currently a Professor and Dean Assistant with the School of Computer and Information, Hefei University of Technology, Hefei, China. His current research interests include pervasive computing and affective computing. He was the recipient of the IEEE Scalcom2009 Excellent Paper Award and NLP-KE2017 Best Paper Award. He is a member of ACM and a senior member of IEEE.
\end{IEEEbiography}

\begin{IEEEbiography}[{\includegraphics[width=1in,height=1.25in,clip,keepaspectratio]{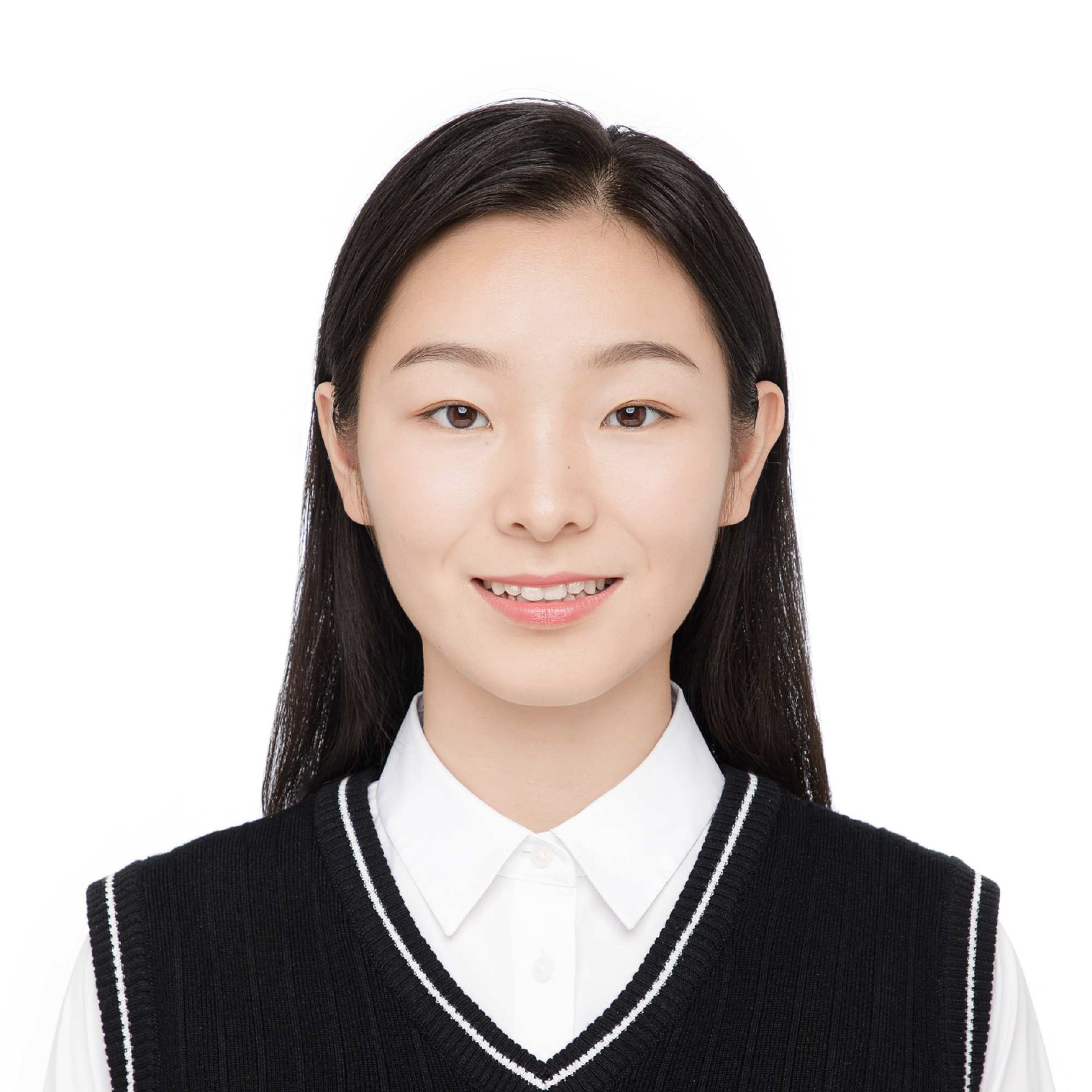}}]{Yantong Wang} received the B.E degree from the Shanghai Normal University in 2016. From 2017 to now, she is a postgraduate student in the Hefei University of Technology. Her research interest includes affective computing and sensorless sensing.
\end{IEEEbiography}

\begin{IEEEbiography}[{\includegraphics[width=1in,height=1.25in,clip,keepaspectratio]{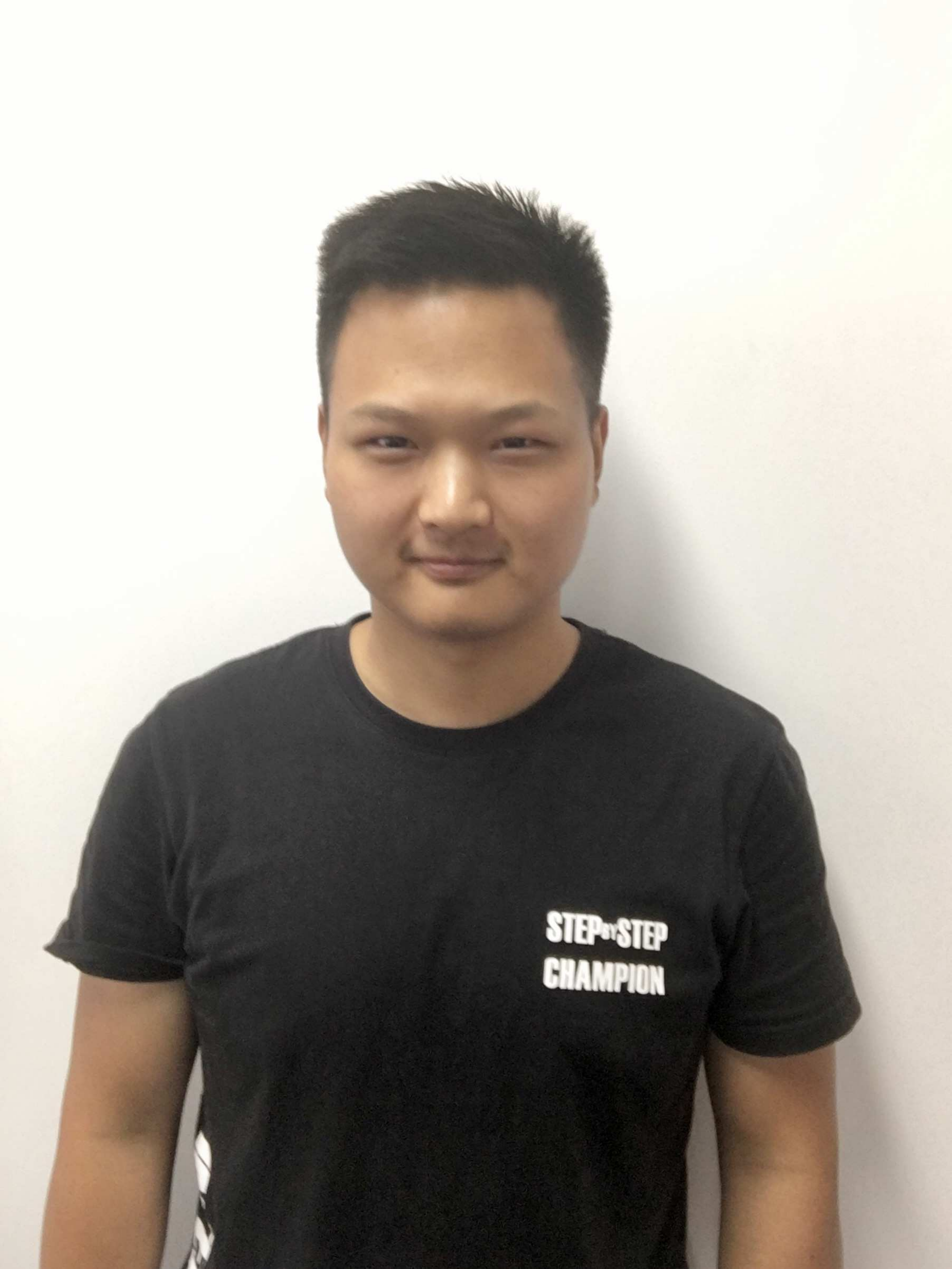}}]{Tao Liu} received the B.E degree from the Anqing Normal University in 2014. From 2016 to now, He is a postgraduate student in the Hefei University of Technology. His research interest includes motion sensing and affective computing.
\end{IEEEbiography}

\begin{IEEEbiography}[{\includegraphics[width=1in,height=1.25in,clip,keepaspectratio]{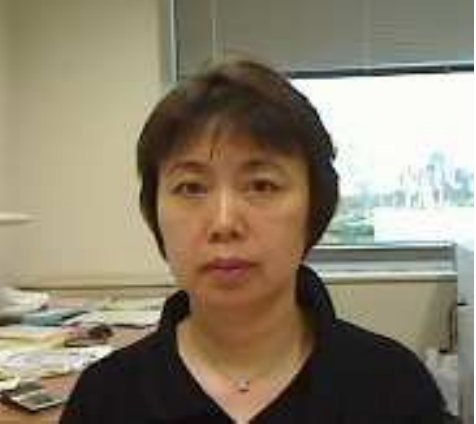}}]
{Yusheng Ji} received B.E., M.E., and D.E. degrees in electrical engineering from the University of Tokyo. She joined the National Center for Science Information Systems, Japan (NACSIS) in 1990. Currently, she is a Professor at the National Institute of Informatics, Japan (NII), and the Graduate University for Advanced Studies (SOKENDAI). She is also appointed as a Visiting Professor at the University of Science and Technology of China (USTC). Her research interests include network architecture, resource management, and performance analysis for quality of service provisioning in wired and wireless communication networks.
\end{IEEEbiography}

\begin{IEEEbiography}[{\includegraphics[width=1in,height=1.25in,clip,keepaspectratio]{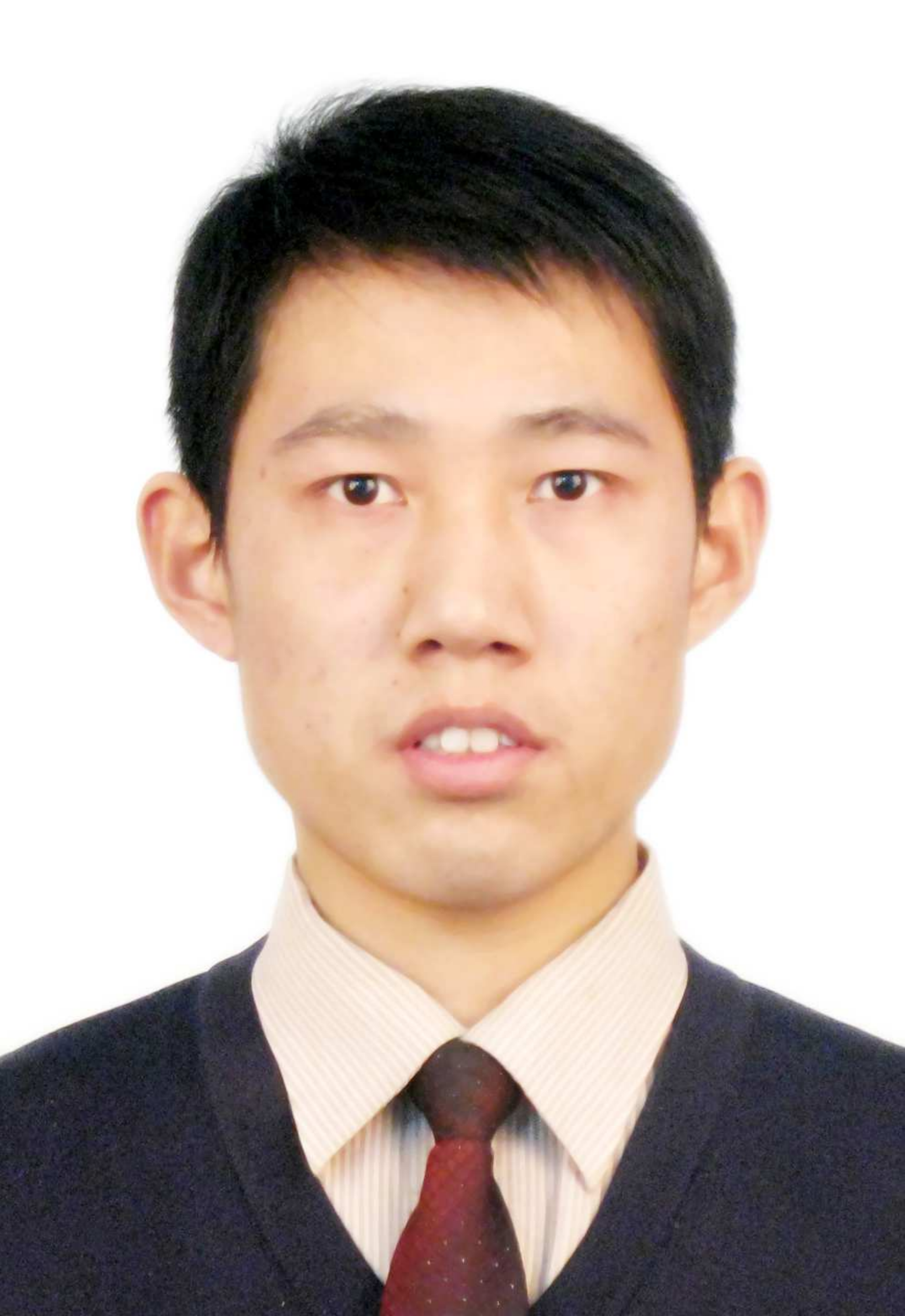}}]{Zhi Liu} (SM‘11-M’14) received the B.E., from the University of Science and Technology of China, China and Ph.D. degree in informatics in National Institute of Informatics. He is currently an Assistant Professor at Shizuoka University. He was a Junior Researcher (Assistant Professor) at Waseda University and a JSPS research fellow in National Institute of Informatics 
 
His research interest includes video network transmission, vehicular networks and mobile edge computing. He was the recipient of the IEEE StreamComm2011 best student paper award, 2015 IEICE Young Researcher Award and ICOIN2018 best paper award. He is and has been a Guest Editor of journals including Wireless Communications and Mobile Computing, Sensors and IEICE Transactions on Information and Systems. He has been serving as the chair for number of international conference and workshops. He is a member of IEEE and IEICE.

\end{IEEEbiography}

\begin{IEEEbiography}[{\includegraphics[width=1in,height=1.25in,clip,keepaspectratio]{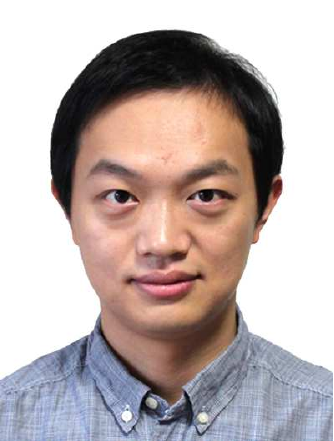}}]{Peng Li} (S'10-M'12) received his BS degree from Huazhong University of Science and Technology, China, in 2007, the MS and PhD degrees from the University of Aizu, Japan, in 2009 and 2012, respectively. He is currently an Associate Professor in the University of Aizu, Japan. His research interests mainly focus on cloud computing, Internet-of-Things, big data systems, as well as related wired and wireless networking problems. He is a member of IEEE.
\end{IEEEbiography}

\begin{IEEEbiography}[{\includegraphics[width=1in,height=1.25in,clip,keepaspectratio]{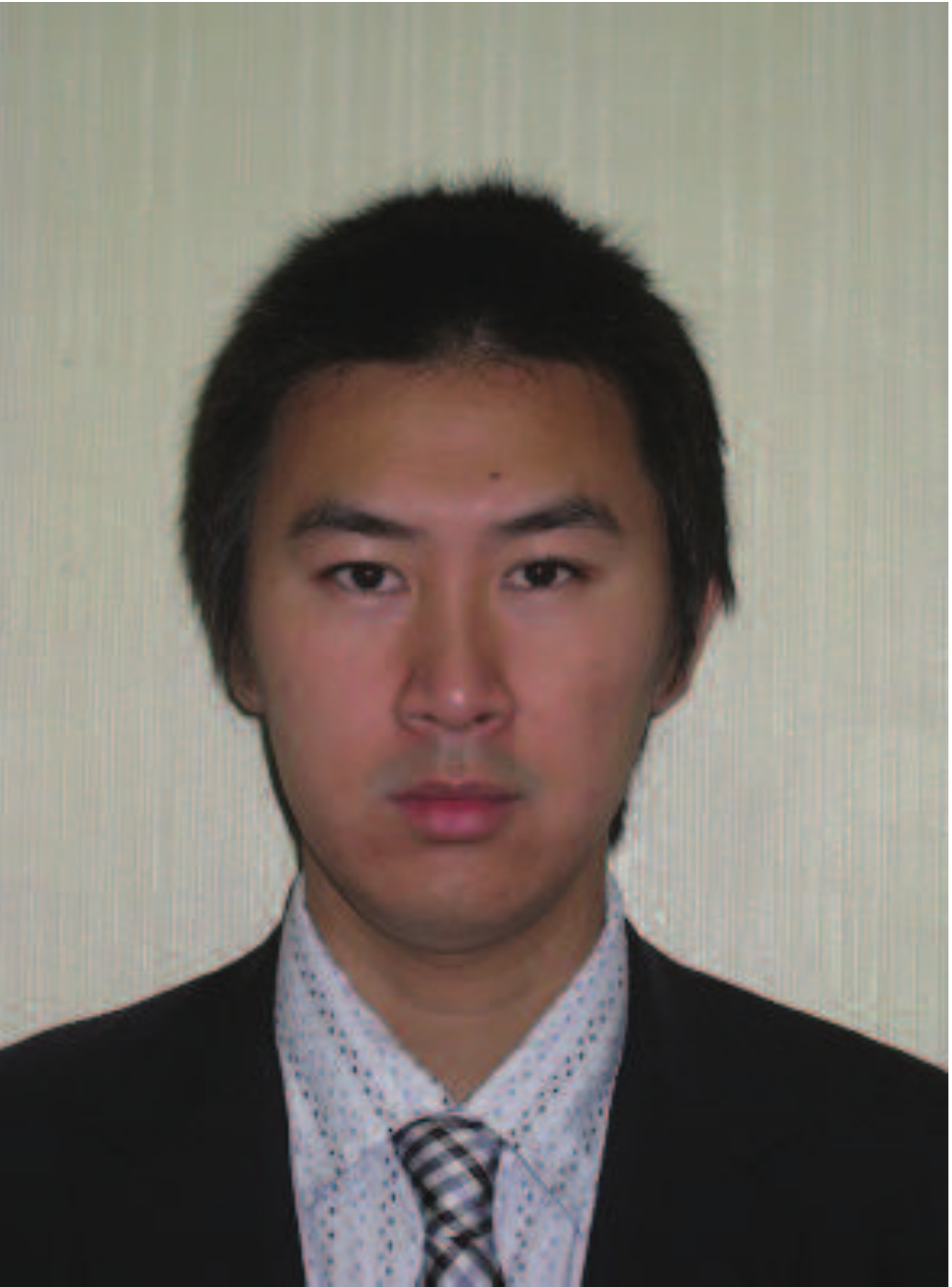}}]{Xiaoyan Wang} received the BE degree from Beihang University, China, and the ME and Ph. D. from the University of Tsukuba, Japan. He is currently working as an assistant professor with the Graduate School of Science and Engineering at Ibaraki University, Japan. Before that, he worked as an assistant professor (by special appointment) at National Institute of Informatics (NII), Japan, from 2013 to 2016. His research interests include networking, wireless communications, cloud computing, big data, security and privacy.
\end{IEEEbiography}

\begin{IEEEbiography}[{\includegraphics[width=1in,height=1.25in,clip,keepaspectratio]{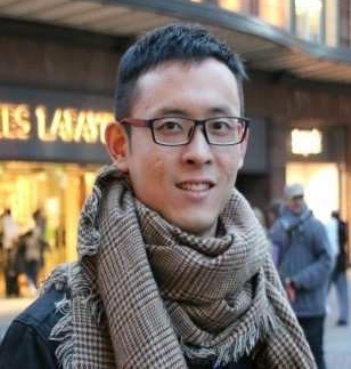}}]{XinAn} is currently an Associate Professor in school of Computer and Information, Hefei University of Technology. He received his bachelor's degree and master's degree in computer science from Shandong University in 2007 and 2010 respectively. From 2010 to 2013, he worked as a Ph.D candidate in INRIA-Grenoble and received his Ph.D degree in Computer Science from Universite de Grenoble in 2013. His research interests focus on the design and control of adaptive embedded systems.
\end{IEEEbiography}

\begin{IEEEbiography}[{\includegraphics[width=1in,height=1.25in,clip,keepaspectratio]{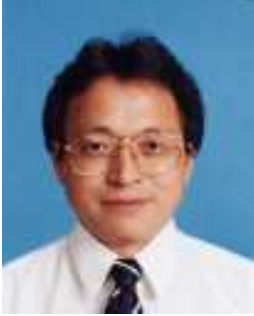}}]{Fuji Ren} received his B.E. and M.E. degrees from Beijing University of Posts and Telecommunications, Beijing, China, in 1982 and 1985, respectively. He received his Ph.D. degree in 1991 from Hokkaido University, Japan. He is a professor in the Faculty of Engineering of the University of Tokushima, Japan. His research interests include information science, artificial intelligence, language understanding and communication, and affective computing. He is a member of IEICE, CAAI, IEEJ, IPSJ, JSAI, AAMT, and a senior member of IEEE. He is a fellow of the Japan Federation of Engineering Societies. He is the president of the International Advanced Information Institute.
\end{IEEEbiography}
\end{document}